\documentclass[12pt,a4paper]{article}

\usepackage[british]{babel}
\usepackage[a4paper,top=2cm,bottom=2cm,left=2.5cm,right=2.5cm,marginparwidth=1.75cm]{geometry}
\usepackage{tikz}
\usetikzlibrary{decorations.pathreplacing}
\usepackage[braket, qm]{qcircuit}      
\usepackage{graphicx}   
\usepackage{cite}

\usepackage{amsmath}
\usepackage{graphicx}
\usepackage[colorlinks=true, allcolors=blue]{hyperref}
\usepackage{hyperref}
\usepackage[title]{appendix}
\usepackage{mathrsfs}
\usepackage{amsfonts}
\usepackage{booktabs} 
\usepackage{caption}  
\usepackage{threeparttable} 
\usepackage{algorithm}
\usepackage{algorithmicx}
\usepackage{algpseudocode}
\usepackage{listings}
\usepackage{enumitem}
\usepackage{chngcntr}
\usepackage{booktabs}
\usepackage{lipsum}
\usepackage{subcaption}
\usepackage{authblk}
\usepackage[T1]{fontenc}    
\usepackage{csquotes}       
\usepackage{diagbox}
\usepackage[numbers]{natbib}

\usepackage{setspace}
\usepackage{titlesec}
\titleformat{\section} 
  {\normalfont\Large\bfseries}{\thesection.}{1em}{}
  
\usepackage{lineno} 

\rightlinenumbers 

\usepackage{float}   
\usepackage{caption} 


\title{Quantum Approximate and Quantum Walk Optimization Approaches to Set Balancing}

\author[1]{Nikhil Kowshik}
\author[1]{Sayan Manna}
\author[2]{Sudebkumar Prasant Pal}

\affil[1]{Department of Metallurgical \& Materials Engineering, IIT Kharagpur}
\affil[2]{Department of Computer Science \& Engineering, IIT Kharagpur}

\begin{document}
\date{}
\maketitle
\begin{abstract}


We explore the application of variational quantum algorithms to the NP-hard set balancing problem, a critical challenge in clinical trial design and experimental scheduling. The problem is mapped to an Ising model, with tailored Quadratic Unconstrained Binary Optimization (QUBO) formulations and cost Hamiltonians expressed in Pauli-Z form. We implement both the Quantum Approximate Optimization Algorithm (QAOA) and the Quantum Walk Optimization Algorithm (QWOA), evaluating them in separate experimental settings. For QAOA, we perform a comparative analysis of six mixer Hamiltonians (X, XY, Full-SWAP, Ring-SWAP, Grover, and Warm-Started), employing scaled-exponential Pauli-string realizations of the mixer unitaries, which yield superior performance over conventional circuit decompositions. Additionally, we introduce a Shannon-entropy-based post-processing technique that refines solutions by maximizing feature-distribution uniformity across partitions. These results underscore the importance of mixer choice and circuit implementation in enhancing QAOA performance for combinatorial optimization.
\end{abstract}



\section{Set Balancing: Introduction}

The problem of splitting a set into two subsets with almost identical properties or attributes but differing in only one of the attributes is a {\it set balancing} problem. It naturally emerges during the design of experiments \cite{fisher1935design,mitzenmacher2005probability} with respect to a collection of attributes. When designing studies pertaining to a group of attributes, it automatically comes to light. Every subject possesses a variety of binary traits or qualities. Being young or old, black or white, tall or small, etc., are examples of attributes. Treatment group ($T$) and control group ($C$) are the two subgroups into which individuals are divided. The goal is to ensure that, for each attribute, the percentage of subjects in ($T$) and ($C$) who share that characteristic is about equal. For example, there should be roughly the same percentage of young people, tall individuals, etc.\ in both groups.  

Consider splitting or partitioning a set $S\subseteq U$ into two subsets, where one subset is ``positive'' in an attribute $A$ and the other subset is ``negative'' on the same attribute $A$. Here, $U$ is the universal set of all items. Let $S_i, 1\leq i\leq m$, be a collection ${\cal S}$ of $m$ arbitrary non-empty proper subsets of $U$. One way of partitioning the subset $S$ is to use a bicoloring $b$ of the universal set $U$, and to project that bicoloring $b$ on $S$, thereby splitting $S$ into two subsets, say $S_b$ and $S_b'$. Similarly, any subset $S_i\in {\cal S}$ can also be partitioned into two sets $S_{i_b}$ and $S_{i_b}'$, for all $1\leq i\leq m$. We consider the notion of balancing the set system or collection ${\cal S}$ by a possible bicoloring $b$ provided the cardinalities of $S_{i_b}$ and $S_{i_b}'$ are more or less equal, for all $1\leq i\leq m$.  

For an arbitrary set system ${\cal S}$, we may not have a perfectly balancing bicoloring $b$. This is closely related to the study of {\it discrepancy theory}, which investigates how evenly elements can be distributed across multiple sets \cite{spencer1985six,chazelle2000discrepancy}. In practice, one often seeks a coloring that minimizes the total imbalance either in the infinity metric or the Euclidean metric. We prefer the Euclidean metric, which enables us to directly invoke the QUBO optimization technique of quadratic objective function optimization, where implement Quantum Approximate Optimization Algorithm (QAOA) \cite{farhi2014qaoa} and Quantum Walk Optimization Algorithm (QWOA) \cite{bennett2021quantum, marsh2020combinatorial}.  


The following is an explanation of the set balancing problem for a universal set \( U \) of \( n \) subjects or elements. The number of subjects in the general population is \( n \). The potential feature or attribute count is \( m \). The subjects are described by \( A \), an \( m \times n \) matrix with entries in \( \{0, 1\} \). Each row denotes a feature or a subset of \( U \), whereas each column denotes a subject or an element of \( U \).  

\[
    a_{ij} = \begin{cases} 
    1, & \text{if subject } j \text{ has feature } i \\
    0, & \text{otherwise}
    \end{cases}
\]

The grouping of subjects is denoted by the vector \( b \), which has \( n \) rows and 1 column. The values in \( b \) are either \( -1 \) or \( 1 \), where \( b_j = 1 \) indicates that subject \( j \) belongs to the treatment group \( T \), and \( b_j = -1 \) indicates that subject \( j \) belongs to the control group \( C \).  

The balancing of a feature is described by \( c = Ab \), which is an \( n \times 1 \) vector. The numeric value of \( c_i \) is the imbalance in feature \( i \). If \( c_i > 0 \), then there are more subjects with feature \( i \) in \( T \), and if \( c_i < 0 \), then there are more subjects with feature \( i \) in \( C \).  

\section{Hamiltonians, Ising Models, and QAOA}

Adiabatic quantum optimization \cite{albash2018adiabatic, farhi2000quantum} has long been studied as a paradigm for approximately solving NP-hard combinatorial optimization problems. The central idea relies on interpolating between two Hamiltonians: a \emph{cost Hamiltonian} $H_C$, whose ground state encodes the solution to the optimization problem, and a \emph{mixer Hamiltonian} $H_M$, whose role is to drive transitions between computational basis states. According to the Adiabatic Theorem \cite{farhi2000quantum}, if a system is initialized in the ground state of $H_M$ and evolved slowly under a time-dependent Hamiltonian that interpolates between $H_M$ and $H_C$, it will remain in its ground state throughout the process and finally encode the solution of the original problem in $H_C$.

A classical Ising Hamiltonian, widely used to model optimization problems \cite{lucas2014ising}, can be written as a quadratic polynomial over $N$ binary spins $s_i \in \{\pm 1\}$:
\begin{equation}
H(s_1, \dots, s_N) = - \sum_{i<j} J_{ij} s_i s_j - \sum_{i=1}^{N} h_i s_i, 
\tag{3.1}
\end{equation}
where $J_{ij}$ are pairwise interaction strengths and $h_i$ are external fields. Its quantum analogue replaces spins with Pauli-$Z$ operators, leading to the cost Hamiltonian
\begin{equation}
H_C = H(\sigma_1^z, \dots, \sigma_N^z), \tag{3.2}
\end{equation}
and a standard choice of mixer Hamiltonian is
\begin{equation}
H_M = -h_0 \sum_{i=1}^{N} \sigma_i^x, \tag{3.3}
\end{equation}
which drives transitions between computational basis states.

The Quantum Approximate Optimization Algorithm (QAOA) \cite{farhi2014qaoa} can be understood as a digitized, Trotterized version of adiabatic quantum evolution. Instead of continuously evolving under a time-dependent Hamiltonian, QAOA alternates between unitaries generated by $H_C$ and $H_M$. Specifically, for a circuit with $p$ layers, the QAOA ansatz is
\begin{equation}
|\psi_p(\boldsymbol{\gamma}, \boldsymbol{\beta})\rangle 
= \prod_{k=1}^{p} e^{-i \beta_k H_M} e^{-i \gamma_k H_C} |+\rangle^{\otimes N}, 
\end{equation}
where the parameters $\boldsymbol{\gamma} = (\gamma_1, \dots, \gamma_p)$ and $\boldsymbol{\beta} = (\beta_1, \dots, \beta_p)$ control the discretized evolution. The initial state $|+\rangle^{\otimes N}$, being the ground state of $H_M$, represents an equal superposition of all possible bitstrings, ensuring that all potential solutions are considered at the outset.

QAOA thus provides a variational hybrid quantum-classical framework: the parameters $(\boldsymbol{\gamma}, \boldsymbol{\beta})$ are optimized using a classical optimizer to minimize the expectation value of $H_C$. As the number of layers $p$ increases, QAOA converges to the continuous-time adiabatic algorithm, though in practice small $p$ often yields good approximations with polynomial resources. Variants and generalizations of QAOA extend the choice of mixer Hamiltonians and problem encodings, broadening its applicability to a wider class of optimization and constraint-satisfaction problems \cite{hadfield2019quantum}.

\begin{figure}[H]
\centering
\includegraphics[width=0.8\linewidth]{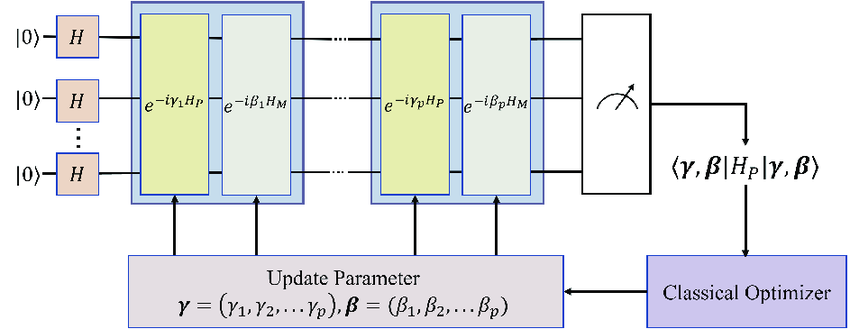}
\caption{\label{fig:1}Alternating layers of \(H_C\) (represented here as \(H_P\))and \(H_M\) form the quantum circuit and are optimized by a classical optimizer. Source for image is \cite{VQACollisionAvoidance2024}}
\end{figure}

These components are schematically illustrated in Figure~\ref{fig:1}.

\section{Set Balancing : Problem Definition}

The \textit{Set Balancing} problem is a fundamental problem in discrepancy theory \cite{chazelle2000discrepancy, mazumdar2009linear} with significant applications in statistical experiment design, computational geometry, and combinatorial optimization. This problem involves finding an optimal bipartitioning of elements such that each subset from a given collection remains as balanced as possible between the two partitions.

\subsection{Matrix Representation}

Let us consider a universe \( U = \{1, 2, \ldots, n\} \) of \( n \) distinct elements, and \( m \) subsets \( A_1, A_2, \ldots, A_m \) of \( U \). Each subset \( A_i \) can be represented as a binary vector in \( \{0,1\}^n \), where a 1 in position \( j \) indicates that the \( j \)-th element of \( U \) is present in subset \( A_i \), and 0 indicates its absence.

These \( m \) binary vectors form an \( m \times n \) matrix \( A \), where each row represents a subset and each column represents an element of \( U \). Specifically, the entry \( a_{ij} \) is 1 if the \( j \)-th element is in the \( i \)-th subset, and 0 otherwise:

\[
A =
\begin{bmatrix}
a_{11} & a_{12} & \cdots & a_{1n} \\
a_{21} & a_{22} & \cdots & a_{2n} \\
\vdots & \vdots & \ddots & \vdots \\
a_{m1} & a_{m2} & \cdots & a_{mn}
\end{bmatrix}.
\]

Since there are \( 2^n \) possible distinct subsets of an \( n \)-element universe, the number of subsets \( m \) is bounded by \( 2^n \). Thus, \( 1 \leq m \leq 2^n \).

\subsection{Problem Formulation}

The Set Balancing problem seeks to find a bipartition of the \( n \) elements of \( U \) into two sets, which we can denote as "red" and "blue" for clarity. This bipartition can be represented by a vector \( b \in \{-1,+1\}^n \), where \( b_j = +1 \) indicates that the \( j \)-th element is colored red, and \( b_j = -1 \) indicates that it is colored blue.

For a subset \( A_i \), represented by the \( i \)-th row of \( A \), the imbalance induced by the bipartition \( b \) is defined as the absolute difference between the number of red elements and the number of blue elements in \( A_i \). This can be expressed as:

\[
\left| N_{+1}(A_i) - N_{-1}(A_i) \right|,
\]

where \( N_{+1}(A_i) \) is the number of elements in \( A_i \) colored red, and \( N_{-1}(A_i) \) is the number of elements colored blue.

It can be shown that this imbalance is equal to the absolute value of the dot product between the \( i \)-th row of \( A \) and the vector \( b \):

\[
| A_i \cdot b | = \left| \sum_{j=1}^{n} a_{ij} \cdot b_j \right|.\tag{4.1}
\]

The objective of the Set Balancing problem is to minimize the maximum imbalance across all subsets, which can be expressed as the infinity norm of the vector \( Ab \):

\[
\| Ab \|_{\infty} = \max_{1 \leq i \leq m} \left| A_i \cdot b \right|.\tag{4.2}
\]

Therefore, the Set Balancing problem can be formulated as the following optimization problem:

\[
\text{minimize} \quad \| Ab \|_{\infty}
\quad \text{subject to} \quad b \in \{-1,+1\}^n.
\]

This problem is known to be NP-hard, meaning that there is no known algorithm that can solve it efficiently for large instances. However, various approximation algorithms and heuristics have been developed to find good, if not optimal, solutions.

\subsection{QUBO Formulation}

In this paper, we fit a Quadratic Unconstrained Binary Optimization (QUBO) \cite{glover2018qubo} model to the Set Balancing problem to solve the problem using quantum methods. Instead of using the \( L_{\infty} \) norm, we use the \( L_2 \) norm to define the objective function. The imbalance vector is given by: \(\mathbf{c} = A \mathbf{b}\), where \( A \) is the \( m \times n \) binary matrix representing attributes, and \( \mathbf{b} \in \{-1,1\}^n \) is the assignment vector indicating group membership. To minimize the overall imbalance, we seek to minimize the squared \( L_2 \) norm of \( \mathbf{c} \):

\[
\min_{\mathbf{b} \in \{-1,1\}^n} \| A \mathbf{b} \|_2^2
\]
Expanding the expression, we get:  
\[
\| A \mathbf{b} \|_2^2 = (A \mathbf{b})^T (A \mathbf{b}) = \mathbf{b}^T A^T A \mathbf{b} = \mathbf{b}^T Q \mathbf{b}, \quad \text{where} \quad Q = A^T A.\tag{4.3}
\]
Thus, the QUBO objective function is:  
\[
\min_{\mathbf{b} \in \{-1,1\}^n} \mathbf{b}^T Q \mathbf{b}.
\]

This represents a standard QUBO formulation, which can be optimized using QAOA.\\
In order to represent the Set Balancing problem in a quantum framework, we define the cost Hamiltonian as follows:

\[
H(\mathbf{b}) = \sum_{i=1}^{m} |c_i|^2  
= \sum_{i=1}^{m} \left| \sum_{j=1}^{n} a_{ij} b_j \right|^2  
= \sum_{i=1}^{m} \left( \sum_{j=1}^{n} a_{ij} b_j \right) \left( \sum_{k=1}^{n} a_{ik} b_k \right).\tag{4.4}
\]

Rewriting in matrix form:

\[
H(\mathbf{b}) = \mathbf{b}^T A^T A \mathbf{b}\tag{4.5}
\]

This cost Hamiltonian captures the total imbalance across all features and serves as the objective function to be minimized in quantum optimization approaches such as QAOA.

\begin{equation*}
\mathbf{A} =
\begin{bmatrix}
    a_{11} & a_{12} & \cdots & a_{1n} \\
    a_{21} & a_{22} & \cdots & a_{2n} \\
    \vdots & \vdots & \ddots & \vdots \\
    a_{m1} & a_{m2} & \cdots & a_{mn}
\end{bmatrix},
\quad
\mathbf{b} =
\begin{bmatrix}
    b_1 \\
    b_2 \\
    \vdots \\
    b_n
\end{bmatrix},
\end{equation*}

The set balancing problem is to find a vector \( \mathbf{b} \) which minimizes the imbalance \( H(\mathbf{b}) \). The value of \( H(\mathbf{b})\) measures the total imbalance across all attributes between the two groups. For each attribute \( i \), the inner summation \( \sum_{j=1}^{n} a_{i,j} \cdot b_j \) represents the net (signed) imbalance for that attribute, and squaring this term penalizes deviations from a balanced state. The outer summation \( \sum_{i=1}^{m} \) accumulates the squares of the signed imbalances for all attributes. If \( H(\mathbf{b}) = 0 \), the set is perfectly balanced for all attributes between the treatment and control groups. Otherwise, \( H(\mathbf{b}) > 0 \) quantifies the degree of imbalance, and the task becomes finding the configuration of \( \mathbf{b} \) that minimizes \( H(\mathbf{b}) \). This function returns the value of the objective function, which we would want to minimize through some optimization process. 

\subsection{Weighted set balancing}

Let us consider the case where each row of the matrix \( \mathbf{A} \) has a scalar weight attached to it, and thus we can say that the individual elements will be multiplied with the values \( K_1, K_2, \dots, K_m \). The balancing of features is described by \( \mathbf{c} = \mathbf{A_k} \mathbf{b} \).

\begin{equation*}
\mathbf{A_k} =
\begin{bmatrix}
    K_1 \cdot a_{11} & K_1 \cdot a_{12} & \cdots & K_1 \cdot a_{1n} \\
    K_2 \cdot a_{21} & K_2 \cdot a_{22} & \cdots & K_2 \cdot a_{2n} \\
    \vdots & \vdots & \ddots & \vdots \\
    K_n \cdot a_{m1} & K_n \cdot a_{m2} & \cdots & K_n \cdot a_{mn}
\end{bmatrix},
\quad
\mathbf{b} =
\begin{bmatrix}
    b_1 \\
    b_2 \\
    \vdots \\
    b_n
\end{bmatrix}
\quad
\mathbf{c} =
\begin{bmatrix}
    c_1 \\
    c_2 \\
    \vdots \\
    c_m
\end{bmatrix}
\end{equation*}

The imbalance of a given partition is defined as

\begin{equation*}
    H(\mathbf{b})  = \sum_{i=1}^{m} |c_i|^2 = \sum_{i=1}^{m} |\sum_{j=1}^{n}  (K_j\times a_{ij} \times b_{j}) |^2\tag{4.6}
\end{equation*}


\subsection{Cost Hamiltonian in Z gate form}
Consider the matrix \( A \):
\[
A = \begin{pmatrix}
a_{11} & a_{12} & \dots & a_{1n} \\
a_{21} & a_{22} & \dots & a_{2n} \\
\vdots & \vdots & \ddots & \vdots \\
a_{m1} & a_{m2} & \dots & a_{mn}
\end{pmatrix}
\]

Consider a random row of \( A \):
\[
A = \begin{pmatrix}
a_{x1} & a_{x2} & \dots & a_{xn} \\
\end{pmatrix}
\]

For the Cost Hamiltonian in the QAOA system, if the values of both \( a_{xi} \) and \( a_{xj} \) are equal to 1, there will be interaction terms involving \( Z_{n-i} \) and \( Z_{n-j} \) (where \( Z \) denotes the Pauli-Z operator). These interaction terms are represented as a tensor product of matrices, where each position corresponds to a specific qubit. The term included in the Cost Hamiltonian corresponding to this row hence takes the form:

\[
I \dots Z_{n-i} \dots Z_{n-j} \dots I = I \otimes I \otimes \dots \otimes Z_{n-i} \otimes \dots \otimes Z_{n-j} \otimes \dots \otimes I
\]

Here:
\begin{itemize}
    \item \( I \) represents the identity matrix.
    \item \( Z_{n-i} \) and \( Z_{n-j} \) are placed at the \( n-i \)-th and \( n-j \)-th positions in the tensor product.
    \item Each \( Z_{n-i} \) or \( Z_{n-j} \) operator corresponds to a qubit where an interaction occurs if \( a_i = a_j = 1 \).
\end{itemize}

Thus, assuming there are \(k\) pair in each row,with \( a_{xi} = a_{xj} = 1 \), with the value of \(k\) changing per row,  the interaction term in the Cost Hamiltonian will be

\[
H_C  = \sum_{p=1}^{m} \sum_{q=1}^{k}  I \otimes I \otimes \dots \otimes Z_{n-i} \otimes \dots \otimes Z_{n-j} \otimes \dots \otimes I.\tag{4.7}
\]

\section{Implementation of QAOA to Set Balancing}
For example, Consider the following A matrix  where \( A \in \{0,1\}^{15 \times 10} \). 
\[
A = \begin{bmatrix}
0 & 1 & 0 & 1 & 1 & 1 & 1 & 0 & 0 & 1 \\
1 & 0 & 1 & 0 & 0 & 0 & 0 & 0 & 0 & 0 \\
1 & 1 & 0 & 0 & 1 & 1 & 1 & 0 & 1 & 0 \\
0 & 1 & 0 & 1 & 1 & 0 & 1 & 0 & 0 & 1 \\
0 & 0 & 1 & 1 & 0 & 1 & 0 & 0 & 0 & 1 \\
0 & 1 & 1 & 0 & 0 & 0 & 1 & 0 & 1 & 0 \\
1 & 1 & 1 & 1 & 1 & 1 & 1 & 1 & 0 & 1 \\
1 & 1 & 0 & 1 & 1 & 0 & 0 & 1 & 0 & 1 \\
1 & 1 & 0 & 0 & 1 & 0 & 1 & 0 & 0 & 1 \\
1 & 1 & 0 & 1 & 1 & 1 & 0 & 1 & 1 & 0 \\
1 & 1 & 0 & 1 & 0 & 0 & 0 & 1 & 0 & 0 \\
1 & 1 & 0 & 0 & 0 & 1 & 1 & 0 & 1 & 0 \\
0 & 0 & 0 & 0 & 1 & 1 & 0 & 1 & 0 & 0 \\
1 & 1 & 0 & 1 & 0 & 1 & 0 & 0 & 0 & 1 \\
0 & 0 & 0 & 1 & 1 & 1 & 0 & 0 & 0 & 1
\end{bmatrix}
\]
Here we are applying standard QAOA method. Following is the qubo formulation of the problem:
\begin{verbatim}
    <class 'qiskit_optimization.problems.quadratic_program.QuadraticProgram'>

Problem name: SetBalancing

Minimize
  36*b_0^2 + 64*b_0*b_1 + 16*b_0*b_2 + 48*b_0*b_3 + 32*b_0*b_4 + 40*b_0*b_5
  + 16*b_0*b_6 + 40*b_0*b_7 + 8*b_0*b_8 + 48*b_0*b_9 + 44*b_1^2 + 16*b_1*b_2
  + 64*b_1*b_3 + 48*b_1*b_4 + 48*b_1*b_5 + 32*b_1*b_6 + 40*b_1*b_7 + 24*b_1*b_8
  + 64*b_1*b_9 + 20*b_2^2 + 24*b_2*b_3 + 16*b_2*b_4 + 8*b_2*b_5 + 16*b_2*b_6
  + 32*b_2*b_8 + 24*b_2*b_9 + 40*b_3^2 + 48*b_3*b_4 + 32*b_3*b_5 + 24*b_3*b_6
  + 24*b_3*b_7 + 32*b_3*b_8 + 72*b_3*b_9 + 32*b_4^2 + 32*b_4*b_5 + 24*b_4*b_6
  + 32*b_4*b_7 + 24*b_4*b_8 + 56*b_4*b_9 + 28*b_5^2 + 16*b_5*b_6 + 24*b_5*b_7
  + 8*b_5*b_8 + 40*b_5*b_9 + 16*b_6^2 + 8*b_6*b_7 + 16*b_6*b_8 + 24*b_6*b_9
  + 24*b_7^2 + 24*b_7*b_9 + 20*b_8^2 + 32*b_8*b_9 + 40*b_9^2 - 192*b_0 - 244*b_1
  - 96*b_2 - 224*b_3 - 188*b_4 - 152*b_5 - 104*b_6 - 120*b_7 - 108*b_8 - 232*b_9
  + 415

Subject to
  No constraints

  Binary variables (10)
    b_0 b_1 b_2 b_3 b_4 b_5 b_6 b_7 b_8 b_9

\end{verbatim}
Now we apply standard QAOA. We are running the program in a simulator. Following is the QAOA configuration details:\\
\begin{itemize}
    \item \textbf{Used classical optimizer:} COBYLA (Constrained Optimization BY Linear Approximations) with a maximum of 500 iterations.
    
    \item \textbf{Depth (Number of QAOA layers, \texttt{reps}):} 3
    
    \item \textbf{Initial values of parameters:}
    \begin{itemize}
        \item The initial values are set as:
        \[
        [\gamma_0, \beta_0, \gamma_1, \beta_1, \gamma_2, \beta_2] = [0.5, 0.5, 0.5, 0.5, 0.5, 0.5]
        \]
        where:
        \begin{itemize}
            \item \(\gamma_i\) are the angles for the cost Hamiltonian at layer \(i\),
            \item \(\beta_i\) are the angles for the mixer Hamiltonian at layer \(i\).
        \end{itemize}
    \end{itemize}
    
    \item \textbf{Number of shots (measurements):} 10000

    \item \textbf{Quantum optimization solver:} 
    \begin{itemize}
        \item MinimumEigenOptimizer from \texttt{qiskit\_optimization.algorithms}, which uses the QAOA instance as the underlying quantum algorithm to solve combinatorial optimization problems.
    \end{itemize}
\end{itemize}

We are getting the following result with the above setting:
\begin{verbatim}
   LOWEST ENERGY(Objective function value) SOLUTIONS (APPROXIMATE BICOLORING)
---------------------------------------------------------------------------
Number of Solutions: 4
Total Probability: 0.58%
Minimum Energy: 11.0000


 Bitstring    Probability             Bicoloring (±1)             Energy 
 ---------    -----------            -----------------           -------

 1011101000       0.25%      [1, -1, 1, 1, 1, -1, 1, -1, -1, -1]     11    

 0111100010       0.11%      [-1, 1, 1, 1, 1, -1, -1, -1, 1, -1]     11    

 1000011101       0.09%      [1, -1, -1, -1, -1, 1, 1, 1, -1, 1]     11    

 0100010111       0.13%      [-1, 1, -1, -1, -1, 1, -1, 1, 1, 1]     11    

\end{verbatim}
Therefore we can see that we are getting 4 bicolorings(2 distinct) corresponding to minimum objective function value.
We are running the program in a quantum simulator, now the probability distribution might change when we run this in real quantum computer. To reduce the statistical noise we increase the number of measurements (shots).
\subsection{Quantum circuit for QAOA}
This is the quantum circuit for standard QAOA for depth = 3 for the previous example
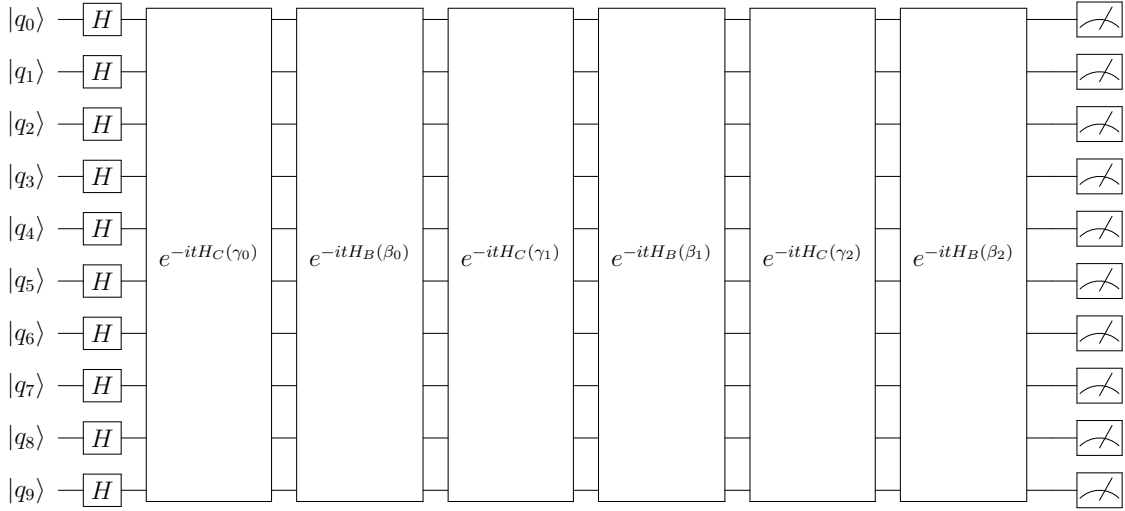
\begin{figure}[H]
    \centering
\scalebox{0.8}{
\Qcircuit @C=1em @R=.7em {
    \lstick{\ket{q_0}} & \gate{H} & \multigate{9}{e^{-i t H_C(\gamma_0)}} & \multigate{9}{e^{-i t H_B(\beta_0)}} & \multigate{9}{e^{-i t H_C(\gamma_1)}} & \multigate{9}{e^{-i t H_B(\beta_1)}} & \multigate{9}{e^{-i t H_C(\gamma_2)}} & \multigate{9}{e^{-i t H_B(\beta_2)}} & \qw & \meter \\
    \lstick{\ket{q_1}} & \gate{H} & \ghost{e^{-i t H_C(\gamma_0)}} & \ghost{e^{-i t H_B(\beta_0)}} & \ghost{e^{-i t H_C(\gamma_1)}} & \ghost{e^{-i t H_B(\beta_1)}} & \ghost{e^{-i t H_C(\gamma_2)}} & \ghost{e^{-i t H_B(\beta_2)}} & \qw & \meter \\
    \lstick{\ket{q_2}} & \gate{H} & \ghost{e^{-i t H_C(\gamma_0)}} & \ghost{e^{-i t H_B(\beta_0)}} & \ghost{e^{-i t H_C(\gamma_1)}} & \ghost{e^{-i t H_B(\beta_1)}} & \ghost{e^{-i t H_C(\gamma_2)}} & \ghost{e^{-i t H_B(\beta_2)}} & \qw & \meter \\
    \lstick{\ket{q_3}} & \gate{H} & \ghost{e^{-i t H_C(\gamma_0)}} & \ghost{e^{-i t H_B(\beta_0)}} & \ghost{e^{-i t H_C(\gamma_1)}} & \ghost{e^{-i t H_B(\beta_1)}} & \ghost{e^{-i t H_C(\gamma_2)}} & \ghost{e^{-i t H_B(\beta_2)}} & \qw & \meter \\
    \lstick{\ket{q_4}} & \gate{H} & \ghost{e^{-i t H_C(\gamma_0)}} & \ghost{e^{-i t H_B(\beta_0)}} & \ghost{e^{-i t H_C(\gamma_1)}} & \ghost{e^{-i t H_B(\beta_1)}} & \ghost{e^{-i t H_C(\gamma_2)}} & \ghost{e^{-i t H_B(\beta_2)}} & \qw & \meter \\
    \lstick{\ket{q_5}} & \gate{H} & \ghost{e^{-i t H_C(\gamma_0)}} & \ghost{e^{-i t H_B(\beta_0)}} & \ghost{e^{-i t H_C(\gamma_1)}} & \ghost{e^{-i t H_B(\beta_1)}} & \ghost{e^{-i t H_C(\gamma_2)}} & \ghost{e^{-i t H_B(\beta_2)}} & \qw & \meter \\
    \lstick{\ket{q_6}} & \gate{H} & \ghost{e^{-i t H_C(\gamma_0)}} & \ghost{e^{-i t H_B(\beta_0)}} & \ghost{e^{-i t H_C(\gamma_1)}} & \ghost{e^{-i t H_B(\beta_1)}} & \ghost{e^{-i t H_C(\gamma_2)}} & \ghost{e^{-i t H_B(\beta_2)}} & \qw & \meter \\
    \lstick{\ket{q_7}} & \gate{H} & \ghost{e^{-i t H_C(\gamma_0)}} & \ghost{e^{-i t H_B(\beta_0)}} & \ghost{e^{-i t H_C(\gamma_1)}} & \ghost{e^{-i t H_B(\beta_1)}} & \ghost{e^{-i t H_C(\gamma_2)}} & \ghost{e^{-i t H_B(\beta_2)}} & \qw & \meter \\
    \lstick{\ket{q_8}} & \gate{H} & \ghost{e^{-i t H_C(\gamma_0)}} & \ghost{e^{-i t H_B(\beta_0)}} & \ghost{e^{-i t H_C(\gamma_1)}} & \ghost{e^{-i t H_B(\beta_1)}} & \ghost{e^{-i t H_C(\gamma_2)}} & \ghost{e^{-i t H_B(\beta_2)}} & \qw & \meter \\
    \lstick{\ket{q_9}} & \gate{H} & \ghost{e^{-i t H_C(\gamma_0)}} & \ghost{e^{-i t H_B(\beta_0)}} & \ghost{e^{-i t H_C(\gamma_1)}} & \ghost{e^{-i t H_B(\beta_1)}} & \ghost{e^{-i t H_C(\gamma_2)}} & \ghost{e^{-i t H_B(\beta_2)}} & \qw & \meter \\
}
}

\caption{Standard QAOA circuit on 10 qubits with depth=3}
\end{figure}
The unitary operator corresponding to the above circuit is
\[
U_{\text{QAOA}} = e^{-i t H_B(\beta_2)} \, e^{-i t H_C(\gamma_2)} \, e^{-i t H_B(\beta_1)} \, e^{-i t H_C(\gamma_1)} \, e^{-i t H_B(\beta_0)} \, e^{-i t H_C(\gamma_0)} H^{\otimes10}\tag{4.8}
\]
Where, $e^{-i t H_B(\beta_i)}$ is Standard mixer Hamiltonian and
\[
e^{-i t H_B(\beta_i)} = \prod_{k=1}^{10} e^{-i t R_{X_{k}}(\beta_i)} 
\]
where,
\[
R_X(\beta) = e^{-i \frac{\beta}{2} X} = \cos\left(\frac{\beta}{2}\right) I - i \sin\left(\frac{\beta}{2}\right) X
\]
and $e^{-i t H_C(\gamma_i)}$ is Cost Hamiltonian,

\subsection{QAOA with Different Mixer Hamiltonians}
In the standard formulation of the QAOA, the choice of mixer Hamiltonian plays a significant role in guiding the quantum state through the solution space. Different mixers induce different types of exploration in the Hilbert space, and their effectiveness can vary depending on the problem structure. Below we describe the six mixer Hamiltonians implemented and studied for the Set Balancing problem.

\subsubsection{X-Mixer (Standard Mixer)}
The X-mixer is the default mixer Hamiltonian used in the original QAOA formulation. It is defined as:
\[
H_M = \sum_{i=1}^n X_i\tag{4.9}
\]
where \( X_i \) is the Pauli-X operator on qubit \( i \). This mixer drives transitions between basis states by flipping qubits and explores the solution space uniformly. The corresponding quantum circuit consists of single-qubit \( R_X \) rotations parameterized by the mixing angle \( \beta \).\\
Quantum circuit for this mixer for $n=4$ is shown below:

\begin{figure}[h]
    \centering
    \scalebox{1.0}{
        \Qcircuit @C=1.0em @R=0.2em @!R {
             \nghost{{q}_{0} :  } & \lstick{{q}_{0} :  } & \gate{\mathrm{R_X}\,(\mathrm{\beta})} & \qw & \qw\\
             \nghost{{q}_{1} :  } & \lstick{{q}_{1} :  } & \gate{\mathrm{R_X}\,(\mathrm{\beta})} & \qw & \qw\\
             \nghost{{q}_{2} :  } & \lstick{{q}_{2} :  } & \gate{\mathrm{R_X}\,(\mathrm{\beta})} & \qw & \qw\\
             \nghost{{q}_{3} :  } & \lstick{{q}_{3} :  } & \gate{\mathrm{R_X}\,(\mathrm{\beta})} & \qw & \qw\\
        }
    }
    \caption{X-Mixer}
    \label{fig:x mixer-circuit}
\end{figure}
Here,
\[
R_X(\beta) = 
\begin{pmatrix}
\cos\left(\frac{\beta}{2}\right) & -i \sin\left(\frac{\beta}{2}\right) \\
-i \sin\left(\frac{\beta}{2}\right) & \cos\left(\frac{\beta}{2}\right)
\end{pmatrix}
\]
The overall unitary operator for the circuit is:
\[
H_{\text{X-mixer}} = R_X(\beta)^{(q_0)} \otimes R_X(\beta)^{(q_1)} \otimes R_X(\beta)^{(q_2)} \otimes R_X(\beta)^{(q_3)}\tag{4.10}
\]
where $R_X(\beta)^{(q_k)}$ acts on qubit $q_k$.

\subsubsection{XY-Mixer}
The XY-mixer includes two-qubit interactions and preserves Hamming weight of the quantum states. It is given by:
\[
H_M = \sum_{\langle i,j \rangle} (X_i X_j + Y_i Y_j)\tag{4.11}
\]
where the summation runs over selected pairs \( \langle i, j \rangle \). This mixer is suitable for problems with fixed Hamming weight constraints and enables entangling interactions that facilitate more complex dynamics than the standard X-mixer.

\begin{figure}[h]
    \centering
    \scalebox{0.8}{
    \Qcircuit @C=1.0em @R=1.0em @!R {
         \nghost{{q}_{0} :  } & \lstick{{q}_{0} :  } & \qw & \qw & \qw & \qw & \multigate{1}{\mathrm{R_{XX}}\,(\,\beta)}_<<<{0} & \multigate{1}{\mathrm{R_{YY}}\,(\,\beta)}_<<<{0} & \qw & \qw \\
         \nghost{{q}_{1} :  } & \lstick{{q}_{1} :  } & \qw & \qw & \multigate{1}{\mathrm{R_{XX}}\,(\,\beta)}_<<<{0} & \multigate{1}{\mathrm{R_{YY}}\,(1.0\,\beta)}_<<<{0} & \ghost{\mathrm{R_{XX}}\,(\,\beta)}_<<<{1} & \ghost{\mathrm{R_{YY}}\,(\,\beta)}_<<<{1} & \qw & \qw \\
         \nghost{{q}_{2} :  } & \lstick{{q}_{2} :  } & \multigate{1}{\mathrm{R_{XX}}\,(\,\beta)}_<<<{0} & \multigate{1}{\mathrm{R_{YY}}\,(\,\beta)}_<<<{0} & \ghost{\mathrm{R_{XX}}\,(\,\beta)}_<<<{1} & \ghost{\mathrm{R_{YY}}\,(\,\beta)}_<<<{1} & \qw & \qw & \qw & \qw \\
         \nghost{{q}_{3} :  } & \lstick{{q}_{3} :  } & \ghost{\mathrm{R_{XX}}\,(\,\beta)}_<<<{1} & \ghost{\mathrm{R_{YY}}\,(\,\beta)}_<<<{1} & \qw & \qw & \qw & \qw & \qw & \qw \\
    }}
    \caption{XY-Mixer for 4 qubits}
    \label{fig:rxx-ryy-circuit}
\end{figure}
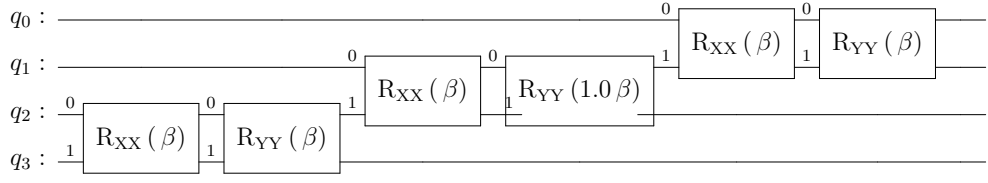
Where, 
\[
R_{XX}(\theta) = e^{-i \frac{\theta}{2} (X \otimes X)} \quad , \quad R_{YY}(\theta) = e^{-i \frac{\theta}{2} (Y \otimes Y)}
\]

\[
R_{XX}(\theta) = 
\begin{pmatrix}
\cos\left(\frac{\theta}{2}\right) & 0 & 0 & -i\sin\left(\frac{\theta}{2}\right) \\
0 & \cos\left(\frac{\theta}{2}\right) & -i\sin\left(\frac{\theta}{2}\right) & 0 \\
0 & -i\sin\left(\frac{\theta}{2}\right) & \cos\left(\frac{\theta}{2}\right) & 0 \\
-i\sin\left(\frac{\theta}{2}\right) & 0 & 0 & \cos\left(\frac{\theta}{2}\right)
\end{pmatrix}
\]

\[
R_{YY}(\theta) = 
\begin{pmatrix}
\cos\left(\frac{\theta}{2}\right) & 0 & 0 & i\sin\left(\frac{\theta}{2}\right) \\
0 & \cos\left(\frac{\theta}{2}\right) & -i\sin\left(\frac{\theta}{2}\right) & 0 \\
0 & -i\sin\left(\frac{\theta}{2}\right) & \cos\left(\frac{\theta}{2}\right) & 0 \\
i\sin\left(\frac{\theta}{2}\right) & 0 & 0 & \cos\left(\frac{\theta}{2}\right)
\end{pmatrix}
\]
The overall unitary operator for the circuit is:
\[
H_{\text{XY-mixer}} = \left( R_{YY}(\beta)_{q_0q_1} \right) \cdot \left( R_{XX}(\beta)_{q_0q_1} \right) \cdot \left( R_{YY}(\beta)_{q_1q_2} \right) \cdot \left( R_{XX}(\beta)_{q_1q_2} \right) \cdot \left( R_{YY}(\beta)_{q_2q_3} \right) \cdot \left( R_{XX}(\beta)_{q_2q_3} \right)\tag{4.12}
\]
where $R_{XX}(\beta)_{q_iq_j}$ mean $R_{XX}$ gate is acting on qubit $q_i, q_j$

\subsubsection{Full-SWAP Mixer}
The Full-SWAP mixer is constructed from full connectivity between all pairs of qubits. It includes terms that effectively perform SWAP operations between qubits:
\[
H_M = \sum_{i < j} \text{SWAP}_{i,j}\tag{4.13}
\]
This mixer allows extensive reshuffling of quantum amplitudes, which can be advantageous in highly symmetric or dense problem instances. The full-SWAP implementation results in highly entangled states.
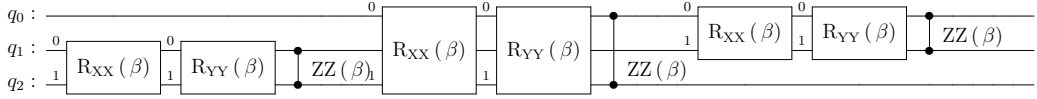
\begin{figure}[h]
    \centering
    \scalebox{0.65}{
    \Qcircuit @C=1.0em @R=0.8em @!R {
         \nghost{{q}_{0} :  } & \lstick{{q}_{0} :  } & \qw & \qw & \qw & \qw & \qw & \qw & \multigate{2}{\mathrm{R_{XX}}\,(\,\beta)}_<<<{0} & \multigate{2}{\mathrm{R_{YY}}\,(\,\beta)}_<<<{0} & \ctrl{2} & \qw & \qw & \qw & \multigate{1}{\mathrm{R_{XX}}\,(\,\beta)}_<<<{0} & \multigate{1}{\mathrm{R_{YY}}\,(\,\beta)}_<<<{0} & \ctrl{1} & \dstick{\hspace{2.0em}\mathrm{ZZ}\,(\,\beta)} \qw & \qw & \qw & \qw & \qw \\
         \nghost{{q}_{1} :  } & \lstick{{q}_{1} :  } & \multigate{1}{\mathrm{R_{XX}}\,(\,\beta)}_<<<{0} & \multigate{1}{\mathrm{R_{YY}}\,(\,\beta)}_<<<{0} & \ctrl{1} & \dstick{\hspace{2.0em}\mathrm{ZZ}\,(\,\beta)} \qw & \qw & \qw & \ghost{\mathrm{R_{XX}}\,(\,\beta)} & \ghost{\mathrm{R_{YY}}\,(\,\beta)} & \qw & \dstick{\hspace{2.0em}\mathrm{ZZ}\,(\,\beta)} \qw & \qw & \qw & \ghost{\mathrm{R_{XX}}\,(\,\beta)}_<<<{1} & \ghost{\mathrm{R_{YY}}\,(\,\beta)}_<<<{1} & \control \qw & \qw & \qw & \qw & \qw & \qw \\
         \nghost{{q}_{2} :  } & \lstick{{q}_{2} :  } & \ghost{\mathrm{R_{XX}}\,(\,\beta)}_<<<{1} & \ghost{\mathrm{R_{YY}}\,(\,\beta)}_<<<{1} & \control \qw & \qw & \qw & \qw & \ghost{\mathrm{R_{XX}}\,(\,\beta)}_<<<{1} & \ghost{\mathrm{R_{YY}}\,(\,\beta)}_<<<{1} & \control \qw & \qw & \qw & \qw & \qw & \qw & \qw & \qw & \qw & \qw & \qw & \qw \\
    }}
    \caption{Full Swap mixer for 3 qubits}
    \label{fig:full-swap-mixer}
\end{figure}
Where, 
\[
ZZ(\theta)=R_{ZZ}(\theta) = e^{-i \frac{\theta}{2} (Z \otimes Z)}
\]
thus, 
\[
ZZ(\theta) = 
\begin{pmatrix}
e^{-i \frac{\theta}{2}} & 0 & 0 & 0 \\
0 & e^{i \frac{\theta}{2}} & 0 & 0 \\
0 & 0 & e^{i \frac{\theta}{2}} & 0 \\
0 & 0 & 0 & e^{-i \frac{\theta}{2}}
\end{pmatrix}
\]
The overall unitary operator for the circuit is:
\[
\begin{aligned}
H_{\text{Full-SWAP}} = \ & ZZ(\beta)_{q_0 q_1} \cdot R_{YY}(\beta)_{q_0 q_1} \cdot R_{XX}(\beta)_{q_0 q_1} \\
& \cdot ZZ(\beta)_{q_0 q_2} \cdot R_{YY}(\beta)_{q_0 q_2} \cdot R_{XX}(\beta)_{q_0 q_2} \\
& \cdot ZZ(\beta)_{q_1 q_2} \cdot R_{YY}(\beta)_{q_1 q_2} \cdot R_{XX}(\beta)_{q_1 q_2}
\end{aligned}\tag{4.14}
\]

\subsubsection{Ring-SWAP Mixer}
The Ring-SWAP mixer is a sparse version of the full-SWAP mixer, in which SWAP operations are performed only between neighboring qubits arranged in a ring topology:
\[
H_M = \sum_{i=1}^{n} \text{SWAP}_{i, (i+1)\mod n}\tag{4.15}
\]
This structure maintains lower circuit complexity while still introducing nontrivial correlations among qubits. It is useful when limited qubit connectivity is a constraint in hardware.
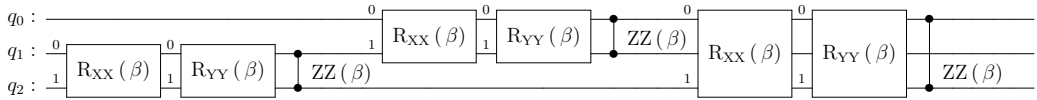
\begin{figure}[h]
    \centering
    \scalebox{0.65}{
    \Qcircuit @C=1.0em @R=0.8em @!R {
         \nghost{{q}_{0} :  } & \lstick{{q}_{0} :  } & \qw & \qw & \qw & \qw & \qw & \qw & \multigate{1}{\mathrm{R_{XX}}\,(\,\beta)}_<<<{0} & \multigate{1}{\mathrm{R_{YY}}\,(\,\beta)}_<<<{0} & \ctrl{1} & \dstick{\hspace{2.0em}\mathrm{ZZ}\,(\,\beta)} \qw & \qw & \qw & \multigate{2}{\mathrm{R_{XX}}\,(\,\beta)}_<<<{0} & \multigate{2}{\mathrm{R_{YY}}\,(\,\beta)}_<<<{0} & \ctrl{2} & \qw & \qw & \qw & \qw & \qw \\
         \nghost{{q}_{1} :  } & \lstick{{q}_{1} :  } & \multigate{1}{\mathrm{R_{XX}}\,(\,\beta)}_<<<{0} & \multigate{1}{\mathrm{R_{YY}}\,(\,\beta)}_<<<{0} & \ctrl{1} & \dstick{\hspace{2.0em}\mathrm{ZZ}\,(\,\beta)} \qw & \qw & \qw & \ghost{\mathrm{R_{XX}}\,(\,\beta)}_<<<{1} & \ghost{\mathrm{R_{YY}}\,(\,\beta)}_<<<{1} & \control \qw & \qw & \qw & \qw & \ghost{\mathrm{R_{XX}}\,(\,\beta)} & \ghost{\mathrm{R_{YY}}\,(\,\beta)} & \qw & \dstick{\hspace{2.0em}\mathrm{ZZ}\,(\,\beta)} \qw & \qw & \qw & \qw & \qw \\
         \nghost{{q}_{2} :  } & \lstick{{q}_{2} :  } & \ghost{\mathrm{R_{XX}}\,(\,\beta)}_<<<{1} & \ghost{\mathrm{R_{YY}}\,(\,\beta)}_<<<{1} & \control \qw & \qw & \qw & \qw & \qw & \qw & \qw & \qw & \qw & \qw & \ghost{\mathrm{R_{XX}}\,(\,\beta)}_<<<{1} & \ghost{\mathrm{R_{YY}}\,(\,\beta)}_<<<{1} & \control \qw & \qw & \qw & \qw & \qw & \qw \\
    }}
    \caption{Ring-SWAP mixer for 3 qubits}
    \label{fig:entangling-circuit-3q}
\end{figure}

The overall unitary operator for the circuit is:
\vspace{-0.5cm}

\begin{align*}
H_{\text{Ring-SWAP}} =\ & ZZ(\beta)_{q_0 q_2} \cdot R_{YY}(\beta)_{q_0 q_2} \cdot R_{XX}(\beta)_{q_0 q_2} \\
& \cdot ZZ(\beta)_{q_0 q_1} \cdot R_{YY}(\beta)_{q_0 q_1} \cdot R_{XX}(\beta)_{q_0 q_1} \\
& \cdot ZZ(\beta)_{q_1 q_2} \cdot R_{YY}(\beta)_{q_1 q_2} \cdot R_{XX}(\beta)_{q_1 q_2}\tag{4.16}
\end{align*}

\subsubsection{Grover Mixer}
The Grover mixer is based on the diffusion operator used in Grover's algorithm. It is problem-specific and generally defined as:
\[
H_M = \mathbb{I} - 2|s\rangle\langle s|\tag{4.17}
\]
where \( |s\rangle \) is the uniform superposition state. The Grover mixer emphasizes amplitude amplification for states with high objective value and can be adapted to perform strong focusing on promising regions of the solution space.
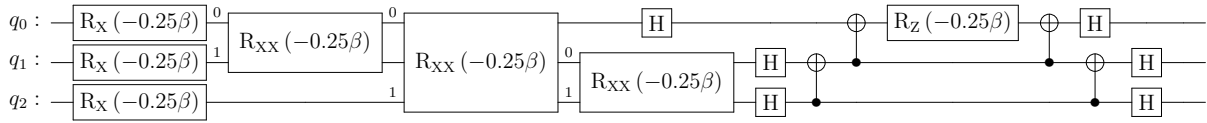
\begin{figure}[h]
    \centering
    \scalebox{0.7}{
    \Qcircuit @C=1.0em @R=0.2em @!R {
        \nghost{{q}_{0} :  } & \lstick{{q}_{0} :  } & \gate{\mathrm{R_X}\,(-0.25\beta)} & \multigate{1}{\mathrm{R_{XX}}\,(-0.25\beta)}_<<<{0} & \multigate{2}{\mathrm{R_{XX}}\,(-0.25\beta)}_<<<{0} & \gate{\mathrm{H}} & \qw & \qw & \targ & \gate{\mathrm{R_Z}\,(-0.25\beta)} & \targ & \gate{\mathrm{H}} & \qw & \qw & \qw \\
        \nghost{{q}_{1} :  } & \lstick{{q}_{1} :  } & \gate{\mathrm{R_X}\,(-0.25\beta)} & \ghost{\mathrm{R_{XX}}\,(-0.25\beta)}_<<<{1} & \ghost{\mathrm{R_{XX}}\,(-0.25\beta)} & \multigate{1}{\mathrm{R_{XX}}\,(-0.25\beta)}_<<<{0} & \gate{\mathrm{H}} & \targ & \ctrl{-1} & \qw & \ctrl{-1} & \targ & \gate{\mathrm{H}} & \qw & \qw \\
        \nghost{{q}_{2} :  } & \lstick{{q}_{2} :  } & \gate{\mathrm{R_X}\,(-0.25\beta)} & \qw & \ghost{\mathrm{R_{XX}}\,(-0.25\beta)}_<<<{1} & \ghost{\mathrm{R_{XX}}\,(-0.25\beta)}_<<<{1} & \gate{\mathrm{H}} & \ctrl{-1} & \qw & \qw & \qw & \ctrl{-1} & \gate{\mathrm{H}} & \qw & \qw \\
    }}
    \caption{Grover mixer for 3 qubits. Global phase: \(-0.375\beta\)}
    \label{fig:grover-mixer-3q}
\end{figure}

The overall unitary operator for the circuit is:
\vspace{-0.5cm}
\begin{align*}
H_{\text{Grover}} =\ & (H_{q_2} \otimes H_{q_1} \otimes H_{q_0}) \cdot \text{CNOT}_{q_2 \rightarrow q_1} \cdot \text{CNOT}_{q_1 \rightarrow q_0} \cdot R_Z(-0.25\beta)_{q_0} \\
& \cdot \text{CNOT}_{q_1 \rightarrow q_0} \cdot \text{CNOT}_{q_2 \rightarrow q_1} \cdot (H_{q_2} \otimes H_{q_1} \otimes H_{q_0}) \\
& \cdot R_{XX}(-0.25\beta)_{q_1 q_2} \cdot R_{XX}(-0.25\beta)_{q_0 q_2} \cdot R_{XX}(-0.25\beta)_{q_0 q_1} \\
& \cdot (R_X(-0.25\beta)_{q_2} \otimes R_X(-0.25\beta)_{q_1} \otimes R_X(-0.25\beta)_{q_0})\tag{4.18}
\end{align*}

\subsubsection{Warm-Started Mixer (Using Classical Relaxation)}
The warm-started QAOA approach uses a classical optimizer (such as CPLEX) to solve a relaxed version of the problem. The solution is then used to initialize the QAOA circuit close to a good classical solution. The mixer Hamiltonian in this case is often still the X-mixer, but the initialization breaks the uniform superposition and leverages prior knowledge to accelerate convergence. This hybrid strategy effectively bridges classical and quantum optimization.

\subsection{Quantum Circuit Model for Scaled Exponential of Pauli Strings}
We implement mixers based on the unitary $e^{i\theta\sigma}$ for any $n$-qubit Pauli string $\sigma$. Sarkar et al. \cite{sarkar2024scalablequantumcircuitsexponential} introduce a scaled-exponential Pauli-string circuit model that realizes $e^{i\theta\sigma}$ using only single-qubit rotations and CNOTs. The key insight is that any Pauli string can be reduced (by conjugation and permutation) to a simple form where only one qubit has an $X$ operator. In particular, any two Pauli strings made of $I$ and $X$ are permutation-similar: a product of CNOT gates (with the last qubit as control) can map $\sigma$ into $I^{\otimes(n-1)}\otimes X$. As a result, the circuit for $e^{i\theta\sigma}$ is implementable on a low-connectivity (star-like) architecture and scales by adding gates for additional qubits. The design uses only Hadamard ($H$), phase ($S$), $R_x$ rotations and CNOTs, without requiring all-to-all qubit coupling.\\
This construction is scalable and hardware-efficient: adding an $(n+1)$th qubit only requires extending the permutation with one additional CNOT and possibly an $H$ or $S$ gate. Since all entangling operations use the last qubit as control/target, the method is compatible with low-connectivity devices.

\subsubsection*{Application of Scaled Exponential method to Mixer Hamiltonians}

Here it concerns the \emph{circuit-level} implementation of mixer unitaries rather than the selection of different mixer Hamiltonians. The conventional X-mixer and other commonly used mixer Hamiltonians (e.g., XY-, SWAP- and Grover-type mixers) are of course constructible by standard gate decompositions. However, we compile the quantum circuits for all mixers considered using the scalable Pauli-string exponential decomposition method described above. Concretely, for any Pauli string \(\sigma\) we implement the elementary unitary $U_M(\beta)=e^{-i\beta\sigma}$
and compose mixer unitaries as exponentials of single or summed Pauli strings using the scalable compilation described above. Importantly, this is a choice of \emph{how} the mixer unitaries are implemented: the Hamiltonians themselves remain the same. In our QAOA simulations for Set Balancing we applied the scaled Pauli-string exponential circuit decomposition to the X-mixer as well as to the other mixers; circuits constructed in this way consistently produced improved approximation ratios and faster convergence compared to the corresponding conventional (e.g., independent $R_x$-based) implementations. Thus the observed performance gains arise from the scaled-exponential circuit decomposition at the implementation level, not from changing the underlying mixer Hamiltonians.

\subsection{Result: QAOA on set Balancing}
\begin{figure}[H]
\centering
\includegraphics[width=0.8\linewidth]{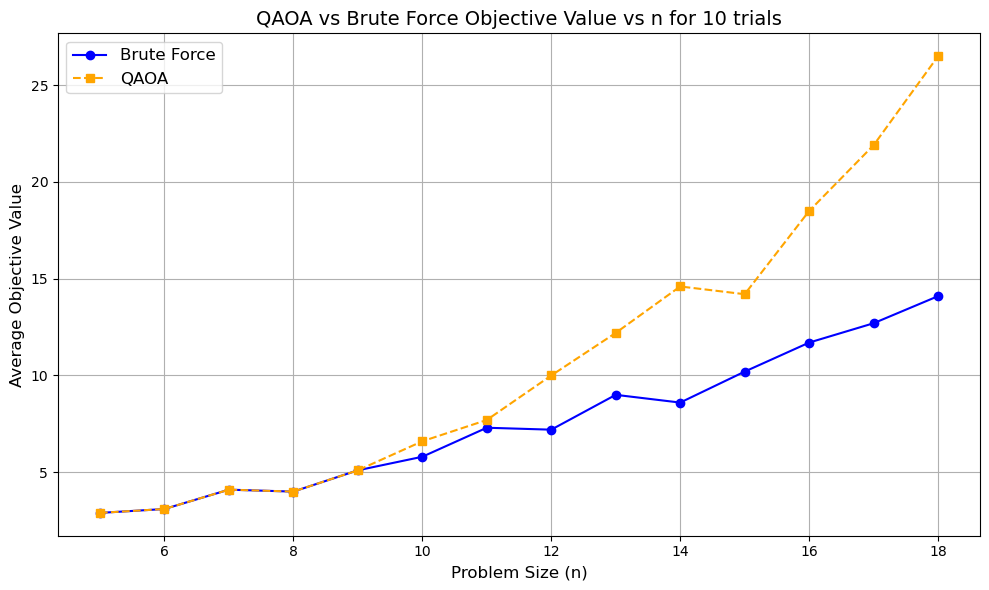}
\caption{\label{fig:result_1}Comparison between brute force and QAOA methods for different matrix sizes}
\end{figure}

We tested the method for graphs of varying sizes, and then compared the ground state energy obtained by the brute force method and QAOA. We performed it on various randomly generated graphs and took an average of the ground state energies of each of the methods, so as to have a fair comparison. We ran each matrix size for \(20\) iterations. For matrix sizes till \(10 \times 10\), the difference between the ground state and QAOA results remained negligible. However, starting from size \( 11 \times 11 \), a noticeable difference began to emerge, with the gap between the ground state and QAOA results continuing to increase for this size and larger matrices. For sizes \( 19 \times 19 \) and \( 20 \times 20 \), this difference became more pronounced, indicating a growing discrepancy between the ground state and QAOA as matrix dimensions increased. This trend suggests that the QAOA method may lose accuracy compared to the ground state brute-force solution as the matrix size grows beyond \( 11 \times 11 \). The comparison is detailed in \ref{fig:result_1}. 
To test the performance of QAOA, we have to consider the average Approximation Ratio with respect to different Mixers. The Approximation Ratio \(\alpha\) is a useful and important metric that can be defined as:

\[
\alpha = \frac{\text{Ground State Objective Value}}{\text{Objective Value Given by QAOA}}
\]

We then tested the performance of different QAOA Mixers and took the average of the approximation ratios for 20 iterations per mixer, each with depth \(p = 5\). The results are shown in the figure, and the most consistent performers are Grover Mixer and SWAP Mixers. It is to be noted that the average \(\alpha\) falls to the range of \(0.7\) are matrix size \(14 \times 14\).

\begin{figure}[H]
\centering
\includegraphics[width=0.7\linewidth]{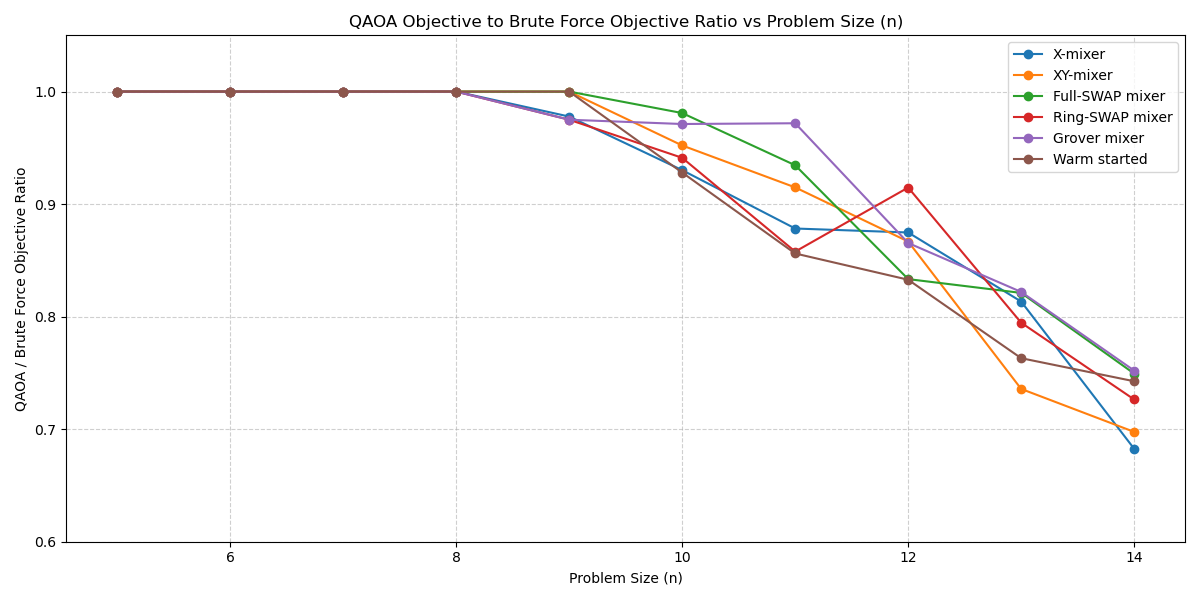}
\caption{\label{fig:result_2}Comparison between various Mixers' performances for the Set Balancing Problem}
\end{figure}

We then tested the performance of the decomposition method applied on different and took the average of the approximation ratios for 20 iterations per mixer, each with only a of depth \(p = 1\). It is here that we can show the power of the Decomposition method, as it outperforms standard QAOA mixers at higher depths (the \(\alpha\) is always in the range \([0.83, 0.95]\). This is even considering that the number of CNOT gates used is lesser or almost equal for depth \(p=1\) as opposed to the total number of CNOT gates used in the standard QAOA method at depth \(p=5\). Even in the decomposition method, Grover and SWAP Mixers still take the lead and have a stunning \(\alpha\) at even large matrix sizes such as \(14 \times 14\). 

\begin{figure}[H]
\centering
\includegraphics[width=0.8\linewidth]{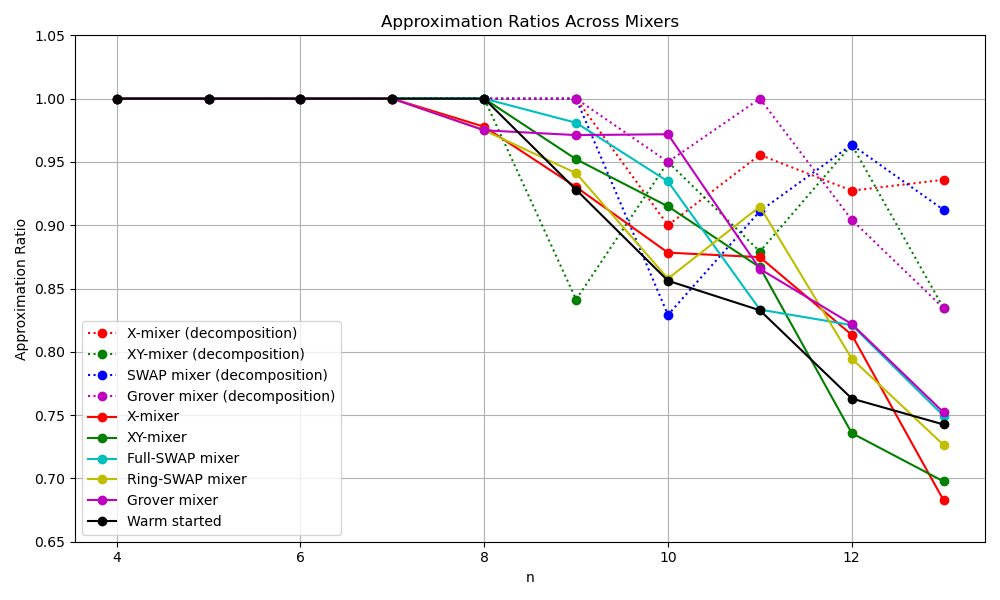}
\caption{\label{fig:result_2}Comparison between various Mixers' performances for the Set Balancing Problem (even decomposition)}
\end{figure}

 \section{QWOA and its Application in Set Balancing}
In this section, we show the implementation of QWOA \cite{wang2018quantum} on set balancing problem.
We consider a mapping \( f: \mathbb{S} \to \mathbb{R} \), which returns a measure of the cost  associated with each possible solution in the solution space \( \mathbb{S} \), where \( \mathbb{S} \) has cardinality \( M \).

We start with a initial quantum system with \(M\) basis states, one for each solution is \(\mathbb{S}\), initialised in an equal superposition, 
\[
| s \rangle = \frac{1}{\sqrt{M}} \sum_{x \in \mathbb{S}} | x \rangle\tag{5.1}
\]

This initial state is then evolved through repeated
application of the quality-dependent phase-shift and
quantum-walk-mixing unitaries. The quality-dependent
phase-shift unitary is given by,
\[
U_Q(\gamma_j) = \exp(-i \gamma_j Q)\tag{5.2}
\]
where \( \gamma_j \in \mathbb{R}\) and \( Q \) is a diagonal operator such that \(Q|x\rangle = f(x)|x\rangle \). This \( U_Q \) applies a phase shift at each node proportional to the cost/quality of the solution at that node, with the proportionality constant given by the parameter \( \gamma_j \).

Next, The quantum-walk-mixing unitary is defined as
\[
U_W(t_j) = \exp(-i t_j \mathcal{L})\tag{5.3}
\]
where \( t_j \geq 0 \), and \(\mathcal{L} \) is the Laplacian matrix of a circulant graph that connects the feasible solutions to the problem. This \(U_W\) can be understood as performing a
quantum walk over the complete graph for time \(t_j\), mixing the
amplitudes across nodes\\
Purposely, we choose a complete graph that
connects computational basis states that correspond to a valid
solution to the problem. Thus, the
Laplacian is deﬁned as
\[
\langle x | \mathcal{L} | y \rangle = 
\begin{cases} 
0 & \text{if } x \text{ or } y \text{ are not solutions} \\ 
M - 1 & \text{if } x = y \\ 
-1 & \text{if } x \neq y 
\end{cases}\tag{5.4}
\]

So, first we apply \(U_Q\) on the initial state \(| s \rangle\), then we apply \(U_W\) on the resultant state. Thus repeating this process for \(r\) times will give us near optimal state. Following the mixing of phase-shifted amplitudes across the nodes of the complete graph, constructive and destructive interference will result in quality-dependent amplitude ampliﬁcation, controlled by the parameters \(\gamma_j\) and \(t_j\).
The application of \( U_Q \) and \( U_W \) is repeated \( r \) times such that the final state of the system is given by

\[
|\gamma, t\rangle = U_{QWOA}( \gamma, t ) | s\rangle  = U_W(t_r) U_Q(\gamma_r) \ldots U_W(t_1) U_Q(\gamma_1) |s\rangle.\tag{5.5}
\]
where \( t = (t_1, t_2, \ldots, t_r) \) and \( \gamma = (\gamma_1, \gamma_2, \ldots, \gamma_r) \)\\
By tuning the parameters \( \gamma \) and \( t \), it is possible to amplify the amplitudes corresponding to low-cost solutions, and therefore increase the probability of a measurement of the system collapsing into a low-cost solution. The process of tuning the parameters is conducted iteratively through the use of a classical optimization algorithm (e.g., Nelder-Mead)which
takes as its objective function the expectation value of the \(Q\)
operator:

\[
c(\gamma, t) = \langle \gamma, t | Q | \gamma, t \rangle\tag{5.6}
\]
\begin{figure}[H]
    \centering
    \includegraphics[width=0.5\linewidth]{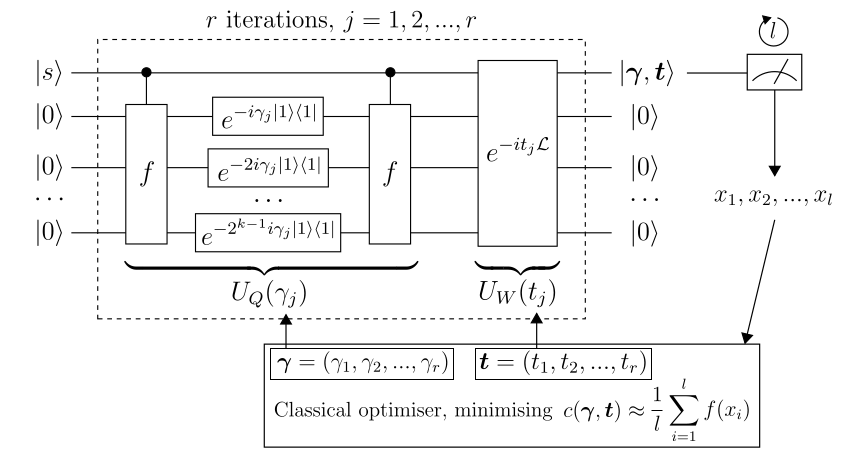} 
    \caption{Schematic diagram of the QWOA circuit paired with a classical optimizer. Adapted from \cite{bennett2021quantum}}
    \label{fig:qwoa_circuit}
\end{figure}
\subsection{Applying QWOA to Set Balancing}
In set balancing problem, the feaible solution space is itself the whole space formed by all possible bicoloring vectors $\mathbf{b} \in \{-1, 1\}^n$. Now, the quality dependent phase shift operator is based on the objective function value corresponding to all possible $b$. Then, quantum walk mixing operator helps to explore the bicoloring space with the help of quantum walk. Finally, after $r$ iterations with classical optimization of parameters, we get the probability distribution of the quantum walker to be at a particular energy state or equivalently on bicoloring$(b)$ state. The value of objective function is calculated similarly from the qubo model of set balancing i.e., $\mathbf{b}^T Q \mathbf{b}$ . Therefore here the walking space of the quantum walker is the space of all possible $b$ vectors.

\subsection{Modified method for QWOA}
Consider the following steps:
\begin{itemize}
    \item Run QAOA for depth \(p =1\), so as to get an approximate solution for the objective value of the objective function. Let this crude solution have objective value equal to \(x\).
    \item Run Grover's Search with a modified Oracle that gets the subset of bicolorings which give the objective value lesser than or equal to the value of \(x\). The modified Oracle is therefore:
    \[
f(b) =
\begin{cases}
1, & \text{if } g(b) \leq x \\
0, & \text{otherwise}
\end{cases}
\]
\item Run QWOA in this reduced subspace to perform a Quantum Random Walk here, where it can find the optimal solution more efficiently 

\end{itemize}

\subsection{Result: QWOA on Set balancing}
Lets consider the following matrix $A$ where $m=n=10$:
\[
A = \begin{bmatrix}
0 & 1 & 1 & 1 & 0 & 1 & 1 & 0 & 1 & 1 \\
1 & 1 & 1 & 0 & 1 & 1 & 0 & 1 & 0 & 1 \\
1 & 0 & 1 & 1 & 0 & 1 & 1 & 1 & 1 & 1 \\
0 & 0 & 0 & 0 & 1 & 1 & 0 & 1 & 0 & 1 \\
0 & 1 & 1 & 1 & 1 & 1 & 0 & 0 & 0 & 1 \\
0 & 1 & 1 & 1 & 1 & 1 & 1 & 0 & 0 & 1 \\
0 & 0 & 0 & 1 & 1 & 0 & 0 & 0 & 0 & 0 \\
0 & 1 & 0 & 1 & 0 & 1 & 1 & 1 & 1 & 0 \\
0 & 1 & 0 & 0 & 0 & 1 & 1 & 0 & 0 & 0 \\
1 & 1 & 1 & 0 & 1 & 0 & 0 & 0 & 1 & 1 \\
\end{bmatrix}
\]
The resultant probability distributions of the quantum waklker to be in an energy state with the increasing depth and at a constant value of number of optimization runs per depth = 4 are given below:
\begin{figure}[H]
    \centering
    \includegraphics[width=0.8\textwidth]{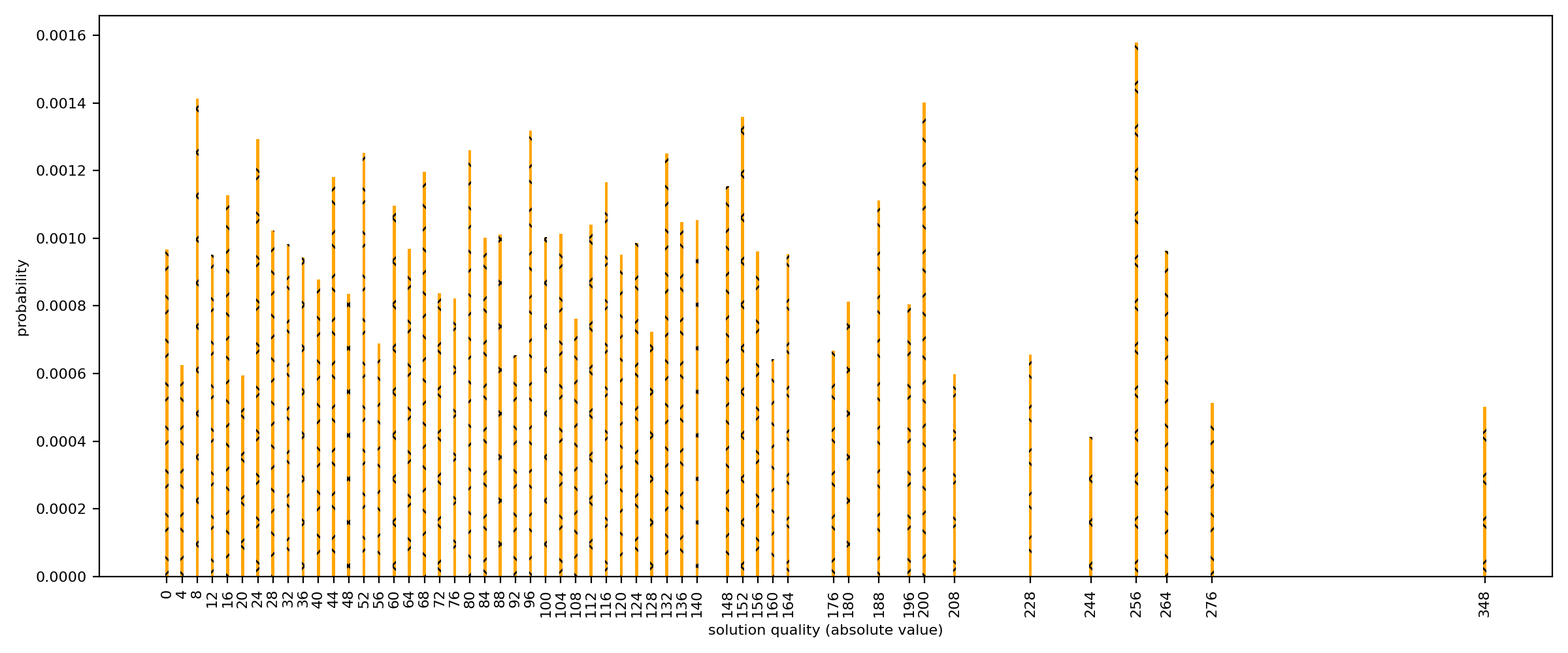}
    \caption{At depth=5}
    \label{fig:qw5_4}
\end{figure}

\begin{figure}[H]
    \centering
    \includegraphics[width=0.8\textwidth]{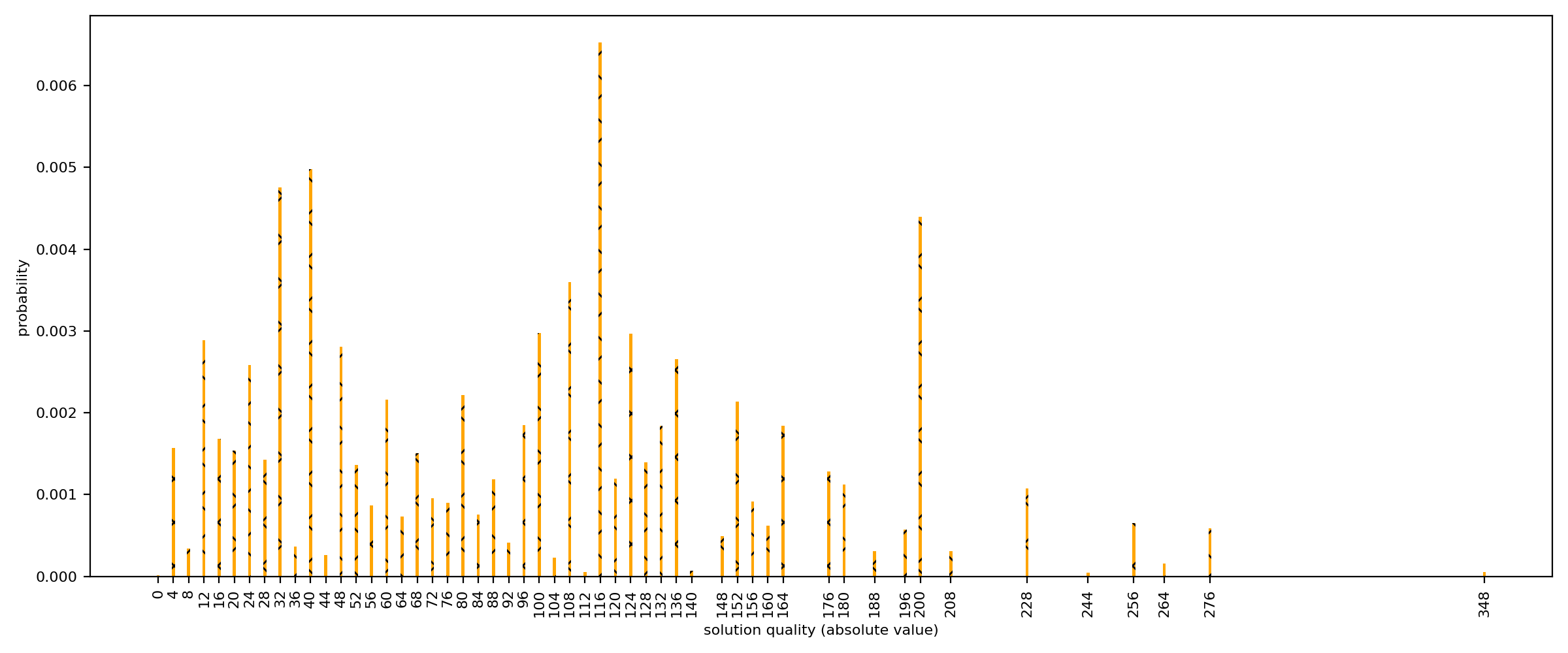}
    \caption{At depth=8}
    \label{fig:qw8_4}
\end{figure}

\begin{figure}[H]
    \centering
    \includegraphics[width=0.8\textwidth]{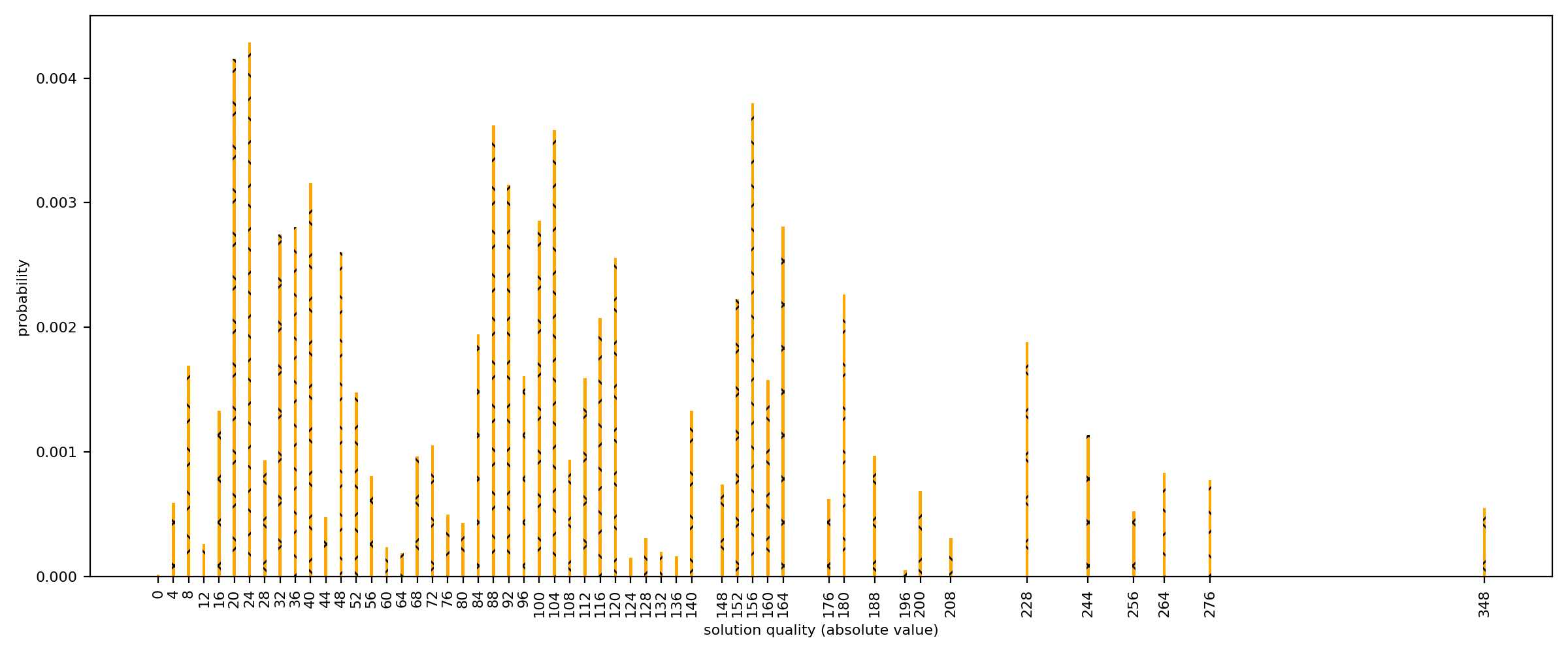}
    \caption{At depth=12}
    \label{fig:qw12_4}
\end{figure}

\begin{figure}[H]
    \centering
    \includegraphics[width=0.8\textwidth]{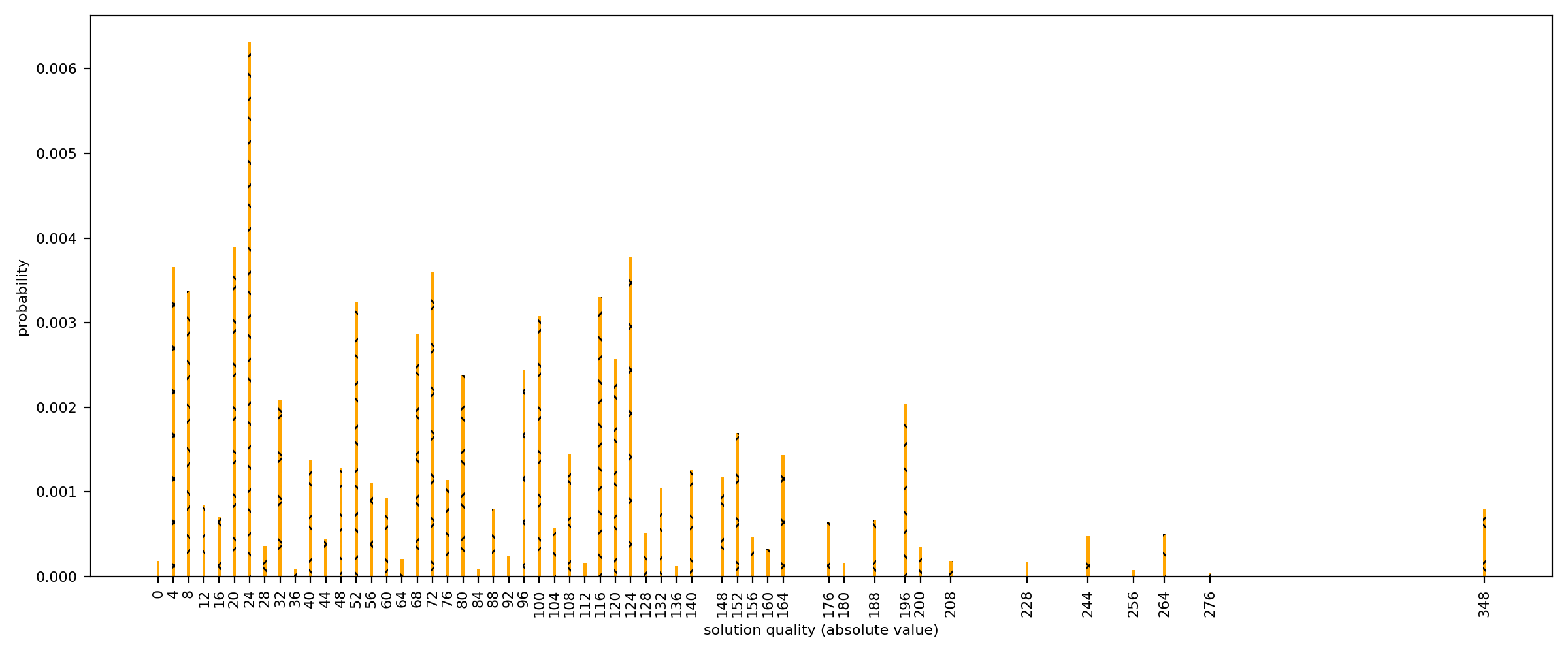}
    \caption{At depth=16}
    \label{fig:qw16_4}
\end{figure}

\begin{figure}[H]
    \centering
    \includegraphics[width=0.8\textwidth]{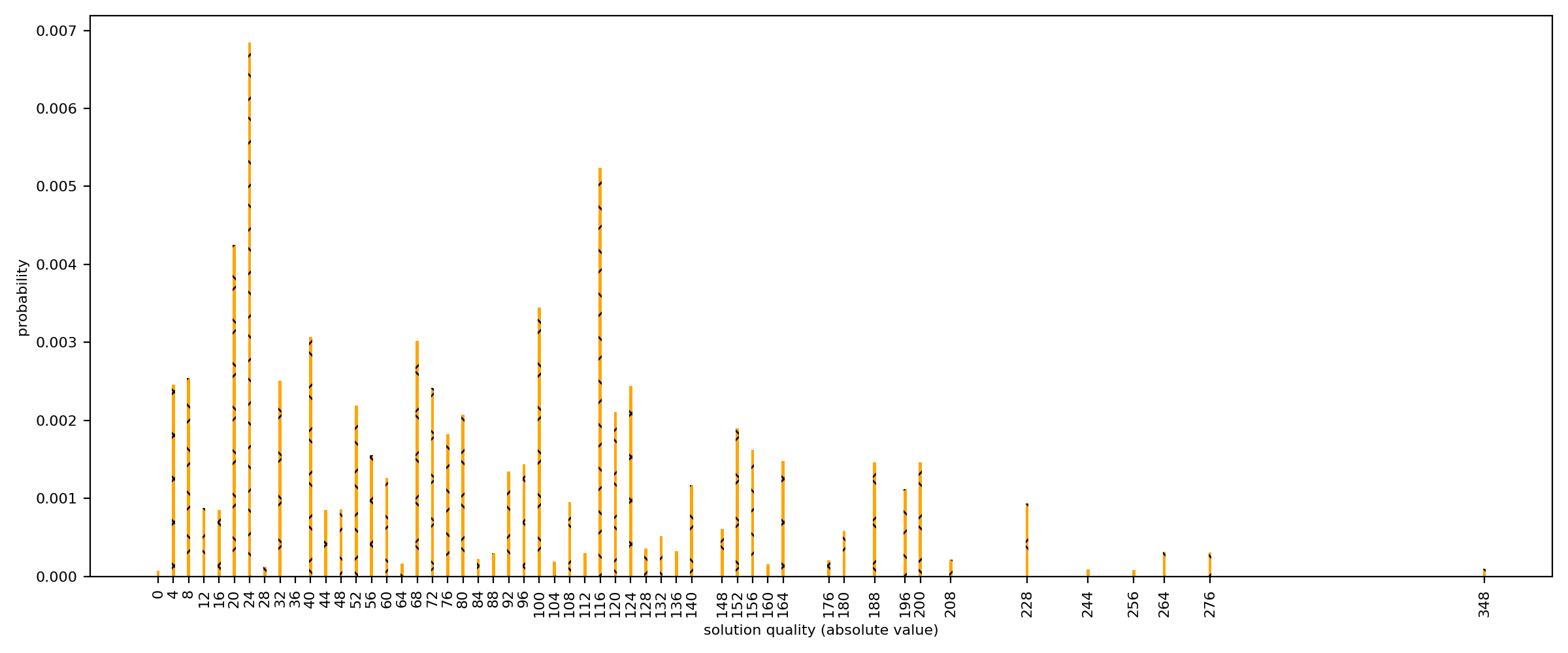}
    \caption{At depth=20}
    \label{fig:qw20_4}
\end{figure}

\begin{figure}[H]
    \centering
    \includegraphics[width=0.8\textwidth]{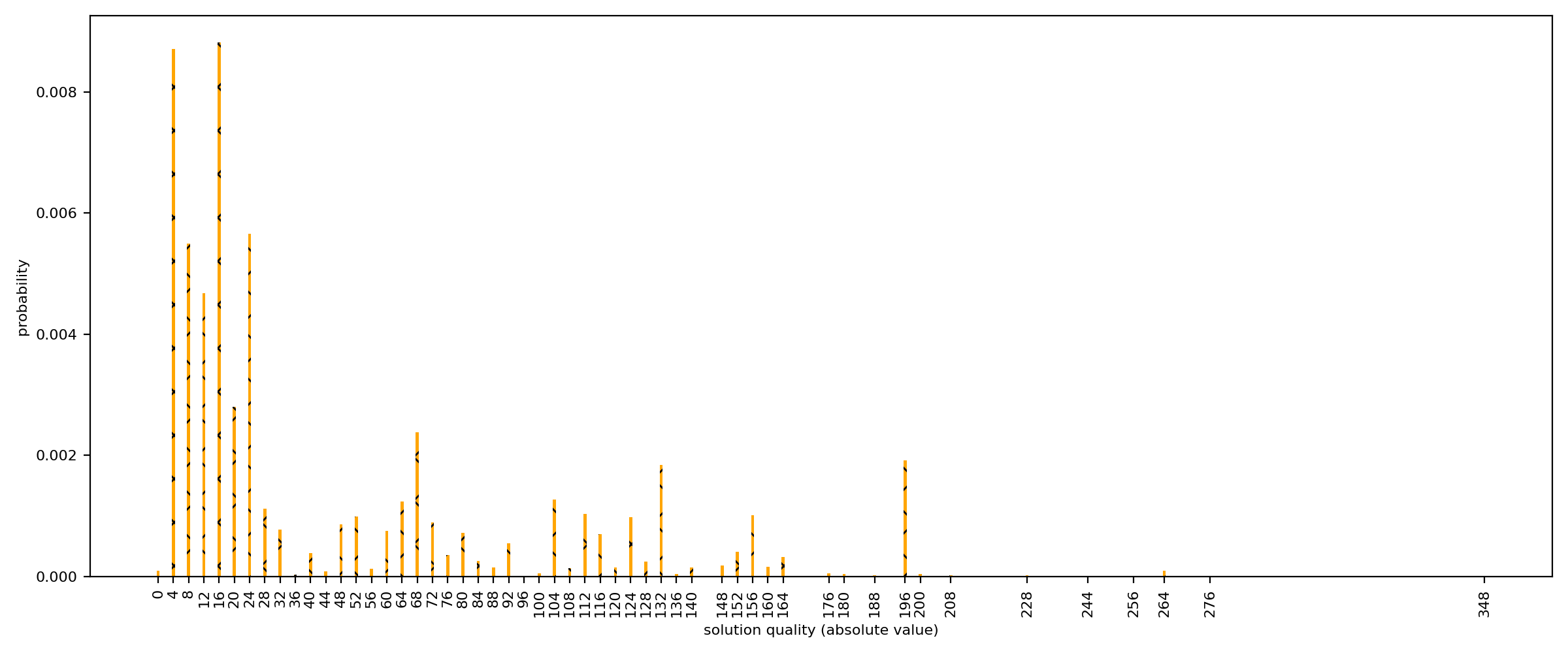}
    \caption{At depth=50}
    \label{fig:qw50_4}
\end{figure}

\begin{figure}[H]
    \centering
    \includegraphics[width=0.8\textwidth]{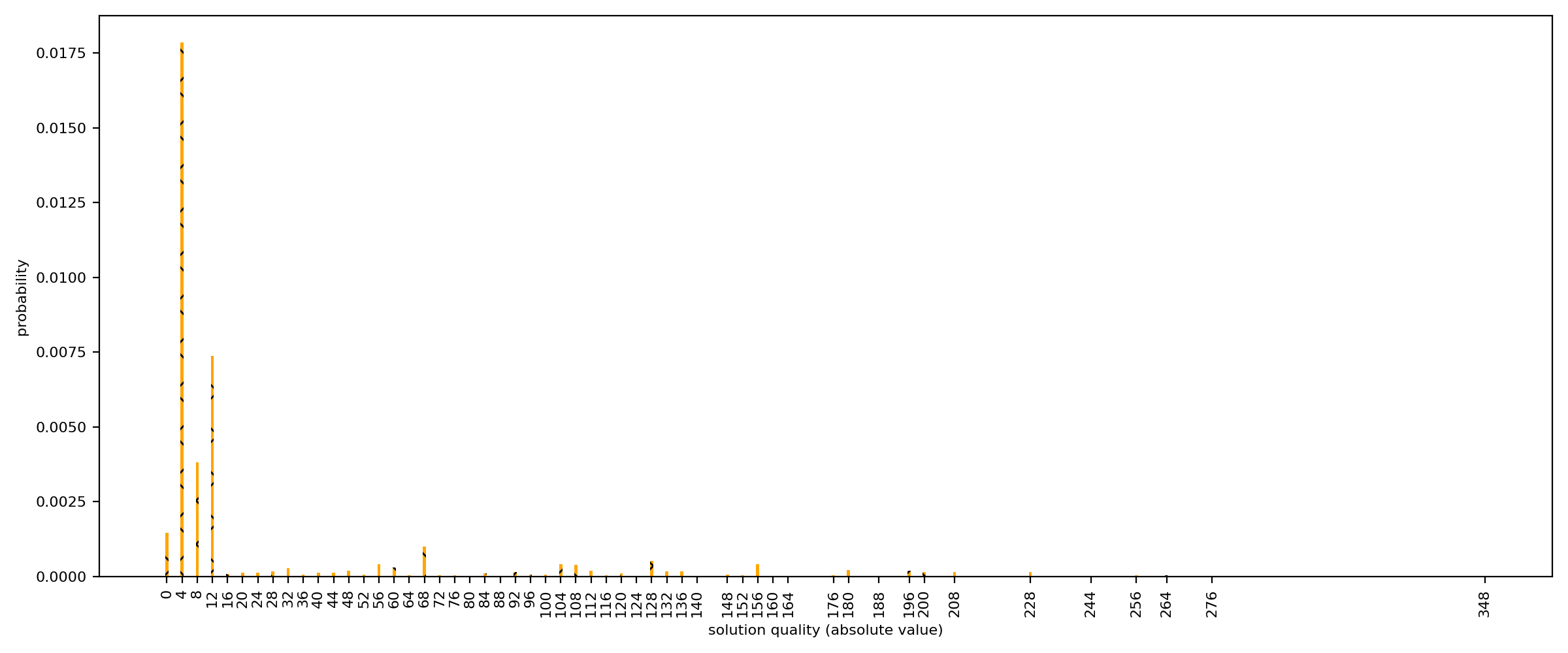}
    \caption{At depth=60}
    \label{fig:qw60_4}
\end{figure}
Therefore with the increawsing number of depths, the probability at being in the minimum energy state is increasing. At depth=60, only nonzero probability exists at energy level=4. Therefore, minimum objective function value is 4. Corresponding bicoloring vector is given by
\( b = [-1,\ -1,\ 1,\ -1,\ 1,\ 1,\ 1,\ -1,\ 1,\ -1] \)
\\
However, from the results we can see that it requires very much higher depth to converge at lower objective function value than QAOA. Therefore we can conclude that QWOA is not efficient in Set Balancing problem as the whole problem space is equal to the quantum walking space or feasible solution space. Therefore QAOA is more efficient in solving set balancing problem than QWOA.

\section{Row-wise post-quantum Shannon entropy optimization}

Among bicolorings with equal approximated (or exact) minimum energy, the bicolourings which provide optimal feature balance distribution across all rows are the ones those maximize total row-wise Shannon entropy and minimize total row-wise entropy difference between partitions.

The procedure for this is:
\begin{enumerate}
    \item Solve the QUBO problem using QAOA to identify the set of optimal/approximate bicolorings $B^*$. We can take the list of the top bicolorings produced by either quantum or classical optimisation.
    \item For each bicoloring $b \in B^*$:
    \begin{itemize}
        \item Partition matrix $A$ into sub-matrices $A_1(b)$ and $A_2(b)$
        \item For a binary sub-matrix $A_j$ resulting from a bicoloring, the row-wise Shannon entropy is defined as:
\begin{align}
H_{\text{row}}(A_j) &= \sum_{i=1}^{m} H_i(A_j)\tag{6.1}
\end{align}
where for each row $i$:
\begin{align}
H_i(A_j) &= -p_i^j \log_2(p_i^j) - (1-p_i^j) \log_2(1-p_i^j) \tag{6.2}\end{align}
Here, $p_i^j$ represent the probabilities of finding a feature (value 1) in row $i$ of the submatrix $A_j$. To select the optimal bicoloring from multiple minimum-energy solutions, we propose a two-stage approach using row-wise entropy.
        \item For each row $i$:
        \begin{itemize}
            \item Calculate probabilities $p_i^1(b)$ and $p_i^2(b)$ of features in each row
            \item Compute Shannon entropies $H_i(A_1(b))$ and $H_i(A_2(b))$
        \end{itemize}
        \item Calculate total row-wise entropy $S_{\text{row}}(b) = H_{\text{row}}(A_1(b)) + H_{\text{row}}(A_2(b))$
    \end{itemize}
    
    \item Identify the bicoloring(s) with minimum total row-wise entropy difference divided by total row-wise entropy:
    \[
    b_{\text{best}}  = b \in B^* =  min \frac{\sum |H_{\text{row}}(A_1(b)) - H_{\text{row}}(A_2(b))|}{\sum H_{\text{row}}(A_1(b)) + H_{\text{row}}(A_2(b))}\tag{6.3}
    \]
\begin{figure}[H]
    \centering
    \includegraphics[width=0.7\textwidth]{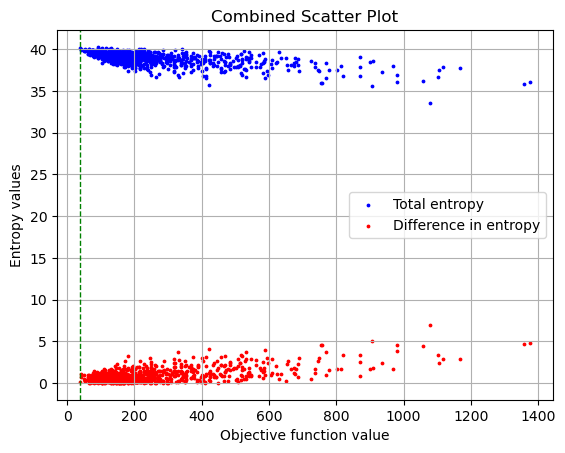}
\end{figure}
The above plot is the plot of Total entropy(E(b)) vs Objective Value(J(b)) [Blue] and Entropy Difference(D(b)) vs Objective Value(J(b)) [red]. The scatter plot is of different bicolorings with probability greater than zero from QAOA. As we can see from the figures Difference in entropy values are coming near zero and higher value of total entropy which is desirable. We can see they are forming almost cluster at high and low values. Now as objective function value is another constraint, therefore we should take a bicoloring which belongs to the leftmost part of these clusters.
\begin{figure}[H]
    \centering
    \includegraphics[width=0.7\textwidth]{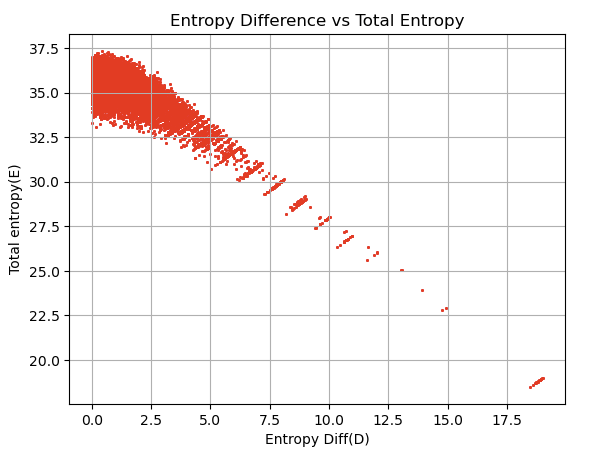}
\end{figure}
The above figure is the plot between Total entropy(E) and Entropy Difference(D) for the all possible bicolorings $(2^n)$ for a particular matrix.

\begin{figure}[H]
    \centering
    \includegraphics[width=0.7\textwidth]{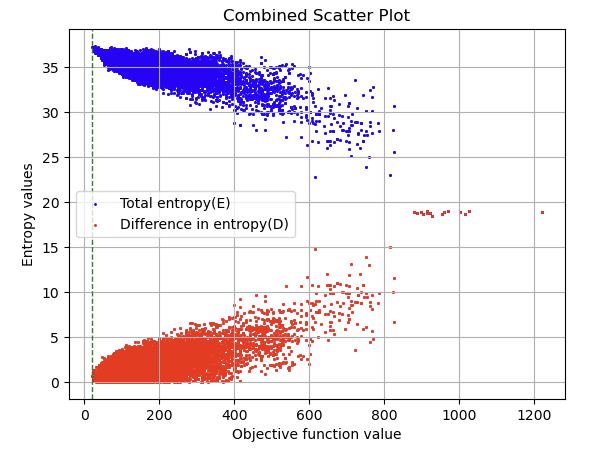}
\end{figure}
The above figure is the plot for all posssible bicolorings $(2^n)$ for a particular matrix.
\subsection{3D Visual Selection of Optimal Bicolorings using Entropy and Objective Measures}

In this section, we propose a unified three-dimensional metric to visualize and identify the best possible bicolorings by integrating three key criteria:

\begin{itemize}
    \item Objective function value, denoted as $J$ (to be minimized),
    \item Total row-wise entropy, denoted as $E$ (to be maximized),
    \item Total row-wise entropy difference, denoted as $D$ (to be minimized).
\end{itemize}

These are formally defined as:
\[
J = \|Ab\|_2^2, \quad D = |H_{\text{row}}(A_1) - H_{\text{row}}(A_2)|, \quad E = H_{\text{row}}(A_1) + H_{\text{row}}(A_2)
\]
where $b$ denotes the bicoloring, $A_1$ and $A_2$ are the resulting partitions of matrix $A$ under $b$, and $H_{\text{row}}(\cdot)$ is the total row-wise Shannon entropy.

\medskip

In the previous section, we considered minimizing the normalized entropy difference, i.e., the ratio \( D/E \). However, to enable a geometric interpretation and multidimensional visualization of the trade-offs among the three objectives, we now reformulate the problem in a three-dimensional Euclidean space. To ensure all objectives are minimized in this formulation, we consider the reciprocal of total entropy, \( 1/E \), instead of maximizing \(E\).

We define the 3D coordinate axes as follows:
\begin{itemize}
    \item $x$-axis: Objective function value, $J$,
    \item $y$-axis: Entropy difference, $D$,
    \item $z$-axis: Reciprocal of total entropy, $1/E$.
\end{itemize}

Since all three quantities are non-negative, we restrict our visualization to the positive octant. We then define the radial distance from the origin in this space as:
\[
r = \sqrt{J^2 + D^2 + \left(\frac{1}{E}\right)^2}\tag{6.4}
\]

The optimal bicoloring(s) are those that minimize this radial distance:
\[
b_{\text{best}} = \arg\min_{b \in B} \sqrt{J(b)^2 + D(b)^2 + \left(\frac{1}{E(b)}\right)^2}\tag{6.5}
\]

To account for near-optimal solutions and allow flexibility in selection, a tolerance threshold $\epsilon$ can be introduced such that all bicolorings satisfying \( r \leq \epsilon \) are retained as viable candidates.

According to the algorithm described in the previous section the formula thus becomes:
\[
b_{\text{best}} = \arg\min_{b \in B} \sqrt{J(b)^2 + \left(\frac{D(b)}{E(b)}\right)^2}\tag{6.6}
\]

\medskip


\end{enumerate}

\section{Set Balancing Interpretations}

There are two Interpretations of Set Balancing in Clinical Trial Design:

\begin{itemize}
\item Interpretation 1: In clinical research, longitudinal studies often require participants to attend multiple follow-up visits over time. Interpretation 1 addresses the challenge of maintaining balanced treatment and control groups across these repeated interactions. Here, the focus is on temporal consistency-ensuring that attendance patterns at various trial stages do not disproportionately favor one study arm. For example, in vaccine trials requiring weekly check-ups, patients might miss appointments due to logistical challenges or personal circumstances. If too many participants from the treatment group skip visits, their cumulative data could become unrepresentative, invalidating comparisons with the control group.

This interpretation models the real-world complexity of participant retention, where researchers must account for "no-show" dynamics while preserving scientific rigor. The goal is to create a treatment allocation strategy that remains robust even when attendance fluctuates unpredictably across trial days.

\item Interpretation 2: This interpretation tackles the foundational challenge of baseline comparability in clinical trials. When testing new therapies, differences in patient characteristics like age, sex, or comorbidities can confound outcomes. For example, if a cancer drug trial accidentally assigns more elderly patients to the control group, age-related mortality rates might obscure the treatment’s true efficacy.

The problem becomes exponentially harder with multiple covariates. A trial balancing diabetes status alone is straightforward, but simultaneously matching sex, BMI, diabetes status, and genetic markers requires sophisticated methods to avoid "covariate collisions"-situations where perfect balance across all factors becomes mathematically impossible.\\
Now suppose we wish to have only one trial, that is we have to find out a suitable subset of $n$ patients, but they must be overall balanced in say $k$ binary attributes like say height (tall/short), weight (heavy/light), sex (male/female), diabetes (yes/no), hypertension (yes/no),, ..., $k$ such attributes. This is a different drug testing problem. These two problems are different and one cannot solve the other directly.\\

Suppose there are 4 patients (\(n=4\)) and we consider 3 binary attributes (\(m=3\)):

\begin{itemize}
    \item Diabetic / Non-diabetic
    \item Hypertension / Non-hypertension
    \item Heart disease / No heart disease
\end{itemize}

The patient conditions are as follows:

\begin{itemize}
    \item Patient 1: Diabetic, does not have hypertension, has heart disease
    \item Patient 2: Non-diabetic, has hypertension, has heart disease
    \item Patient 3: Diabetic, has hypertension, does not have heart disease
    \item Patient 4: Non-diabetic, does not have hypertension, does not have heart disease
\end{itemize}

Now, let \( S_1 \) denote the diabetic status, where a diabetic patient is represented by \(1\) and a non-diabetic patient is represented by \(0\):

\[
S_1 = [1, 0, 1, 0]
\]

Similarly, let \( S_2 \) denote the hypertension status:

\[
S_2 = [0, 1, 1, 0]
\]

And let \( S_3 \) denote the heart disease status:

\[
S_3 = [1, 1, 0, 0]
\]

These three sets can be represented by a matrix \( A \) of size \( 3 \times 4 \), where each row corresponds to an attribute and each column corresponds to a patient:

\[
A = 
\begin{bmatrix}
1 & 0 & 1 & 0 \\
0 & 1 & 1 & 0 \\
1 & 1 & 0 & 0
\end{bmatrix}
\]

Our goal is to divide the 4 patients into 2 groups such that both groups have a roughly balanced distribution of diseases. This ensures that, in the trial, the real drug is given to one group and the placebo is given to the other group for testing.

This problem is analogous to the bi-coloring problem. One possible bi-coloring is:

\[
b = [1, 1, -1, -1]
\]

where \( 1 \) denotes that a patient belongs to one group and \( -1 \) denotes that a patient belongs to the other group. According to this division:

\begin{itemize}
    \item Group 1: Patient 1, Patient 2
    \item Group 2: Patient 3, Patient 4
\end{itemize}

\end{itemize}

\section{Conclusion}

In this work, we have explored the application of variational quantum algorithms to the Set Balancing problem, a fundamental NP-hard problem in discrepancy theory and experimental design. By formulating the problem as a Quadratic Unconstrained Binary Optimization (QUBO) instance, we implemented the Quantum Approximate Optimization Algorithm (QAOA) with multiple mixer Hamiltonians and compared their performance. Additionally, we discussed the Quantum Walk Optimization Algorithm (QWOA) as an alternative variational approach. Our results demonstrate that the choice and structure of the mixer Hamiltonian, particularly when using the scaled exponential of Pauli strings, plays a crucial role in improving the approximation quality and convergence of QAOA. This study highlights the potential of quantum variational algorithms for combinatorial optimization problems and provides insights into designing effective mixers for enhanced performance. Future work may focus on extending these techniques to larger instances, exploring hybrid quantum-classical strategies, and investigating further generalizations of mixer Hamiltonians for other classes of NP-hard problems.

\section*{Acknowledgments}
We would like to thank Sabyasachi Chakrobarty for his valuable assistance in implementing the code for the scaled exponential decomposition of Pauli strings, which significantly contributed to the development of this work.

\bibliography{sample}
\bibliographystyle{plain}



\end{document}